\documentclass[journal]{IEEEtran}
\usepackage[obeyFinal]{easy-todo}
\usepackage{svg}
\usepackage{url}
\usepackage{import}
\usepackage{soul}
\usepackage{booktabs}
\usepackage{threeparttable, tablefootnote}
\usepackage{amsmath}
\usepackage{graphicx}
\usepackage{tikz}
\usetikzlibrary{circuits,circuits.logic}
\usetikzlibrary{circuits.logic.US}
\usepackage{tabularx}
\usepackage{float}

\begin{document}

\title{A CMOL-Like Memristor-CMOS Neuromorphic Chip-Core Demonstrating Stochastic Binary STDP}

\author{L. A. Camuñas-Mesa$^{\dagger}$, E. Vianello$^{*}$, C. Reita$^{*}$, T. Serrano-Gotarredona$^{\dagger}$, and B. Linares-Barranco$^{\dagger}$

\thanks{$\dagger$ Instituto de Microelectr\'{o}nica de Sevilla (IMSE-CNM), CSIC and Universidad de Sevilla, Seville, Spain. 
$*$ CEA-LETI, Grenoble, France. E-mail of corresponding author: \{bernabe@imse-cnm.csic.es\}}
\thanks{This work was partly funded by EU H2020 grants 687299 “NeuRAM3” and 871371 “MeM-Scales” and by grant PID2019-105556GB-C31 (NANOMIND) from Spain’s Ministry of Science and Innovation, with support from the European Regional Development Fund. LCM was funded by the VI PPIT through the Universidad de Sevilla.}
}

\markboth{Journal of \LaTeX\ Class Files,~Vol.~XX, No.~XX, May~2022}%
{Shell \MakeLowercase{\textit{et al.}}: Bare Demo of IEEEtran.cls for IEEE Journals}

\maketitle

\begin{abstract}
The advent of nanoscale memristors raised hopes of being able to build CMOL (\underline{C}MOS/nanowire/mo\underline{l}ecular) type ultra-dense in-memory-computing circuit architectures.
In CMOL, nanoscale memristors would be fabricated at the intersection of nanowires. The CMOL concept can be exploited in neuromorphic hardware by fabricating lower density neurons on CMOS and placing massive analog synaptic connectivity with nanowire and nanoscale-memristor fabric post-fabricated on top. 
However, technical problems have hindered such developments for presently available reliable commercial monolithic CMOS-memristor technologies.
On one hand, each memristor needs a MOS selector transistor in series to guarantee forming and programming operations in large arrays. This results in compound MOS-memristor synapses (called 1T1R) which are no longer synapses at the crossing of nanowires. On the other hand, memristors do not yet constitute highly reliable, stable analog memories for massive analog-weight synapses with gradual learning.
Here we demonstrate a pseudo-CMOL monolithic chip core that circumvents the two technical problems mentioned above by: (a) exploiting a CMOL-like geometrical chip layout technique to improve density despite the 1T1R limitation, and (b) exploiting a binary weight stochastic Spike-Timing-Dependent-Plasticity (STDP) learning rule that takes advantage of the more reliable binary memory capability of the memristors used. Experimental results are provided for a spiking neural network (SNN) CMOL-core with 64 input neurons, 64 output neurons and 4096 1T1R synapses, fabricated in 130nm CMOS with 200nm-sized Ti/HfOx/TiN memristors on top. The CMOL-core uses query-driven event read-out, which allows for memristor variability insensitive computations. Experimental system-level demonstrations are provided for plain template matching tasks, as well as regularized stochastic binary STDP feature-extraction learning, obtaining perfect recognition in hardware for a 4-letter recognition experiment. 
 
\end{abstract}


\section{Introduction}
\IEEEPARstart{O}{ver} the last fifty years, microchip technology has been shaped by an admirable, aggressive trend towards miniaturization known as Moore’s Law \cite{Mack,Moore}. At the time of writing, Intel is already fabricating its new Loihi-2 neuromorphic chip in its Intel-4 node (7nm) technology \cite{Intel} and has a roadmap for upscaling it to 18A (1.8nm) silicon technology by 2025 \cite{Anandtech}. However, the silicon crystal structure has a unit lattice of 0.543nm with a nearest Si atom distance of 0.235nm \cite{Kittel}. This severely limits Moore’s law in terms of the viability of simply continuing to miniaturize Si based devices. In the context of molecular electronics \cite{Villaume}, Likharev proposed the concept of CMOL (CMOS + molecular electronics) \cite{Likharev, Strukov2005}, by which nano-scale molecular devices could be combined with CMOS technology, exploiting the third dimension and thereby boosting overall device density. When the first link between resistive thin-film switches and memristors was reported in 2008 \cite{Strukov2008}, many researchers had great expectations because these devices could potentially be sandwiched at the intersection of nano-wires, implementing very high density non-volatile analog memories massively interconnected with lower-density CMOS neurons \cite{Chua, Strukov2008, Prezioso2015}. Hopes were also raised that CMOL-type hardware structures could be exploited to fabricate massively interconnected neuromorphic systems capable of natural on-line learning \cite{Linares2009, Zamarreno, Linares2018}. 

Reality quickly showed, however, that memristors suffer from several technical limitations, which have so far hindered the commercial development of true CMOL chips for both neuromorphic and other types of applications. In this work, we employed a hybrid CMOS-Memristor technology \cite{Valentian} which uses filamentary HfOx memristors. This type of memristor requires an initial step in which filaments are formed by applying a relatively high voltage (in the range of 4-5V) while limiting the maximum current flowing through the device. To do this, each memristor requires an in-series “selector” device, which in this technology is an nMOS transistor. The combined series-connected nMOS-memristor structure is typically referred to as a “1T1R” memory element or, in the context of neuromorphic computing and engineering, as “1T1R synapses”.

Another technical limitation of currently available filamentary HfOx memristors is their poor capability for programming and holding analog values robustly and reliably. 
Some progress in this respect is being made recently \cite{Wang2016,ElisaArxiv2022}, but the analog programming requires slow iterative procedures because of the inherent stochastic behavior and relaxations observed. 
The practical recommendation is to use them as binary memory elements by setting them to either a high resistance state (HRS), with a typical resistance in the range of $100k\Omega$ or above, or a low resistance state (LRS), with a typical resistance in the range of $10-20k\Omega$. These memristors also show strong inter-device mismatches in their HRS/LRS values, and high write/erase cycle-to-cycle variability for the same device.

Consequently, circuit and system designers need to circumvent these technical limitations by developing electronic circuit techniques and architectures compatible with such limitations. Here we demonstrate, for the first time, a CMOL-like monolithic neuromorphic core fabricated in low-power 130nm CMOS-memristor technology. On one hand, we propose a geometrical layout arrangement technique that allows 1T1R synapses and neurons to share area within a unit tile, thus approaching the original CMOL concept. On the other hand, we exploit a previously reported STDP (spike-timing-dependent plasticity) learning rule variant \cite{Yousefzadeh} which relies exclusively on binary weights by using stochasticity to control full-range rather than gradual weight updates. Thus, we demonstrate a physical implementation of a memristor-based binarized Spiking Neural Network with an unsupervised STDP-like learning method, which can be an alternative or complementary to supervised backpropagation binary neural networks~\cite{Eshraghian}.

The paper is structured as follows. Next Section introduces the CMOL concept and how we have adapted it to the present RRAM technology constraints. Section III descibes the stochastic binary STDP learning rule used. Section IV describes the implemented CMOL-core chip. Section V provides extensive experimental results, and finally Section VI provides the conclusions and future outlooks. Section VII includes an appendix with a brief illustration of a more generic synaptic crossbar.

\section{Pseudo-CMOL memristor crossbar}

\begin{figure}
\centering
\includegraphics[width=1.0in]{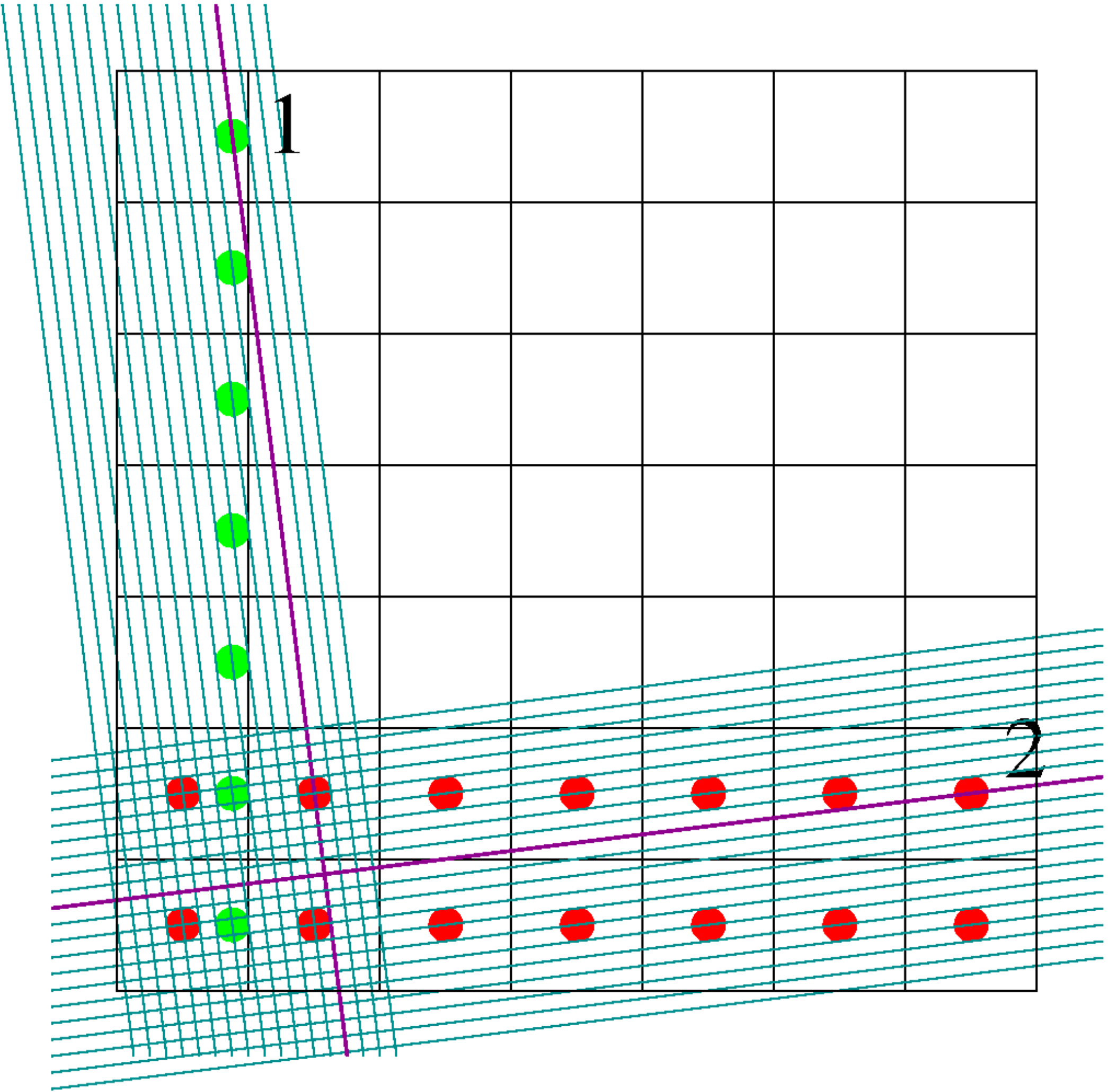}
\caption{Illustration of the CMOL approach with an NxN array of 2D cells, each with one input and one output, all fully-connected.}
\label{fig:Fig1}
\end{figure}

The CMOL approach or architecture was proposed in 2005 by Likharev \cite{Likharev} in the context of molecular electronic devices. The basic idea is illustrated in Fig. \ref{fig:Fig1}, where an NxN 2D array of cells lies under a slightly tilted nanowire crossbar.

Each cell in the array has one input terminal (for example, the red dots in Fig. \ref{fig:Fig1}) and one output terminal (for example, the green dots in Fig. \ref{fig:Fig1}). Each output terminal (green) of each cell connects to a slightly tilted vertical nanowire. Each input terminal (red) of each cell connects to a slightly tilted horizontal nanowire. Each vertical nanowire connects only to one output (green) terminal, and each horizontal nanowire connects only to one input (red) terminal. The tilted angles are set in such a way that this is guaranteed. At the crossing of each vertical and horizontal nanowire there is a synaptic device. 
The scheme represented in Fig. \ref{fig:Fig1} therefore shows a system with $N\times N = N^2$ pre-synaptic circuits fully connected to $N\times N = N^2$ post-synaptic circuits using $N^2 \times N^2 = N^4$ memristors.



\begin{figure}
\centering
\includegraphics[width=3.5in]{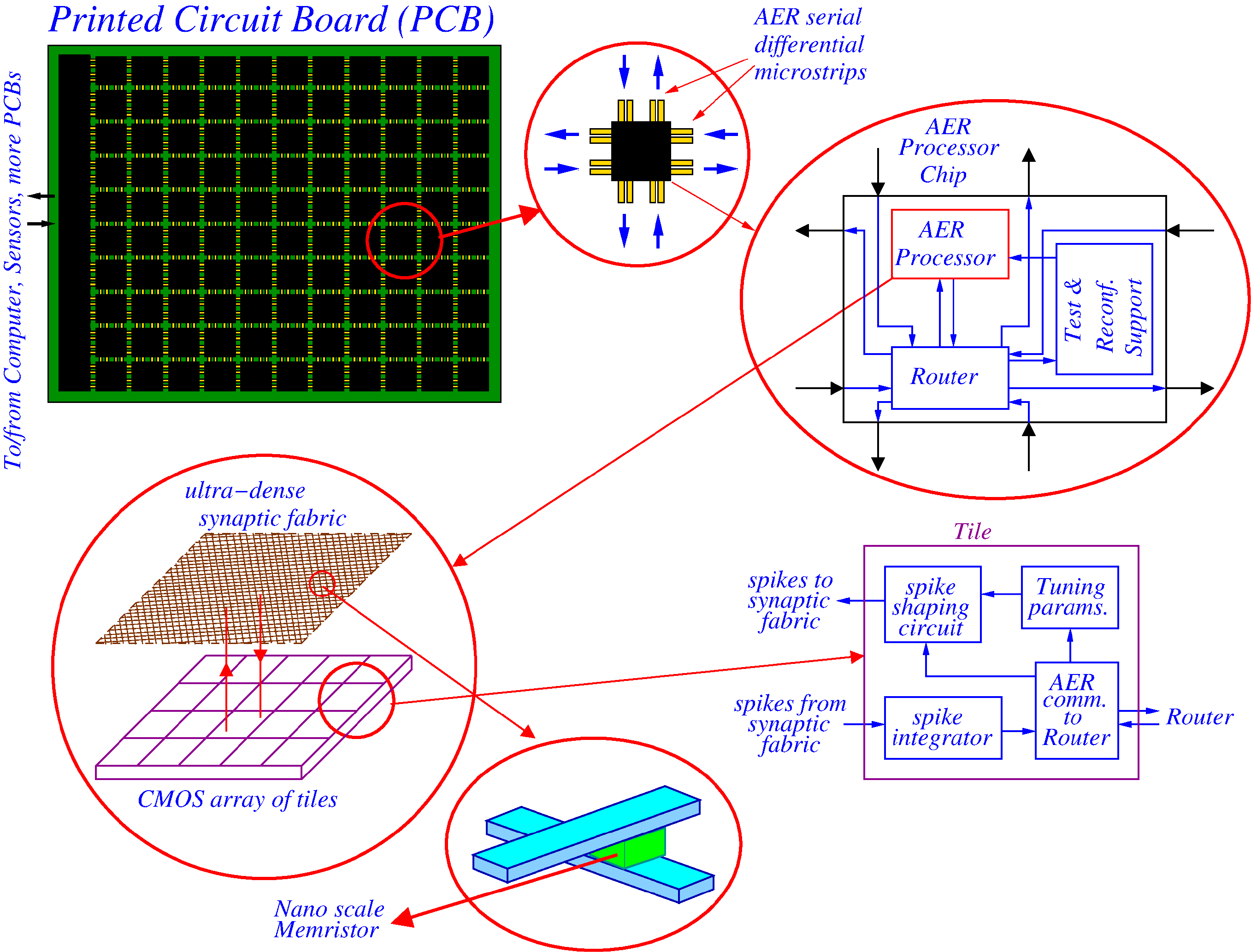}
\caption{Illustration of CMOL arrangement with AER communication.}
\label{fig:Fig2}
\end{figure}

In the case of a Spiking Neural Network (SNN) System,
the cells in Fig. \ref{fig:Fig1} would be receiving and sending spike events through Address-Event-Representation (AER) communication channels, for example compact serial AER-links~\cite{AER_serial,AER_serial_Spinn}. AER assigns to each neuron an address or ID that identifies it. Every time a neuron generates a spike, its address (or ID) is communicated to its destinations through this communication network, which can be intra-chip, inter-chip, inter-PCB, etc. High-flexibility programmable inter-connectivity, either between the internal layers or between other layers allocated to other CMOL crossbars, could thus be configured through the use of AER mappers and routers. 
Fig.~\ref{fig:Fig2} illustrates a multi-chip system assembled on a PCB (printed circuit board), where chips communicate through AER with their neighbors in the same PCB or outside the PCB. Each chip contains an array (or arrays) of tiles massively interconnected with a nano-scale memristive synaptic fabric. Each tile would contain one or more neurons plus additional communication and configuration circuitry. 

\begin{figure}
\centering
\includegraphics[width=3.4in]{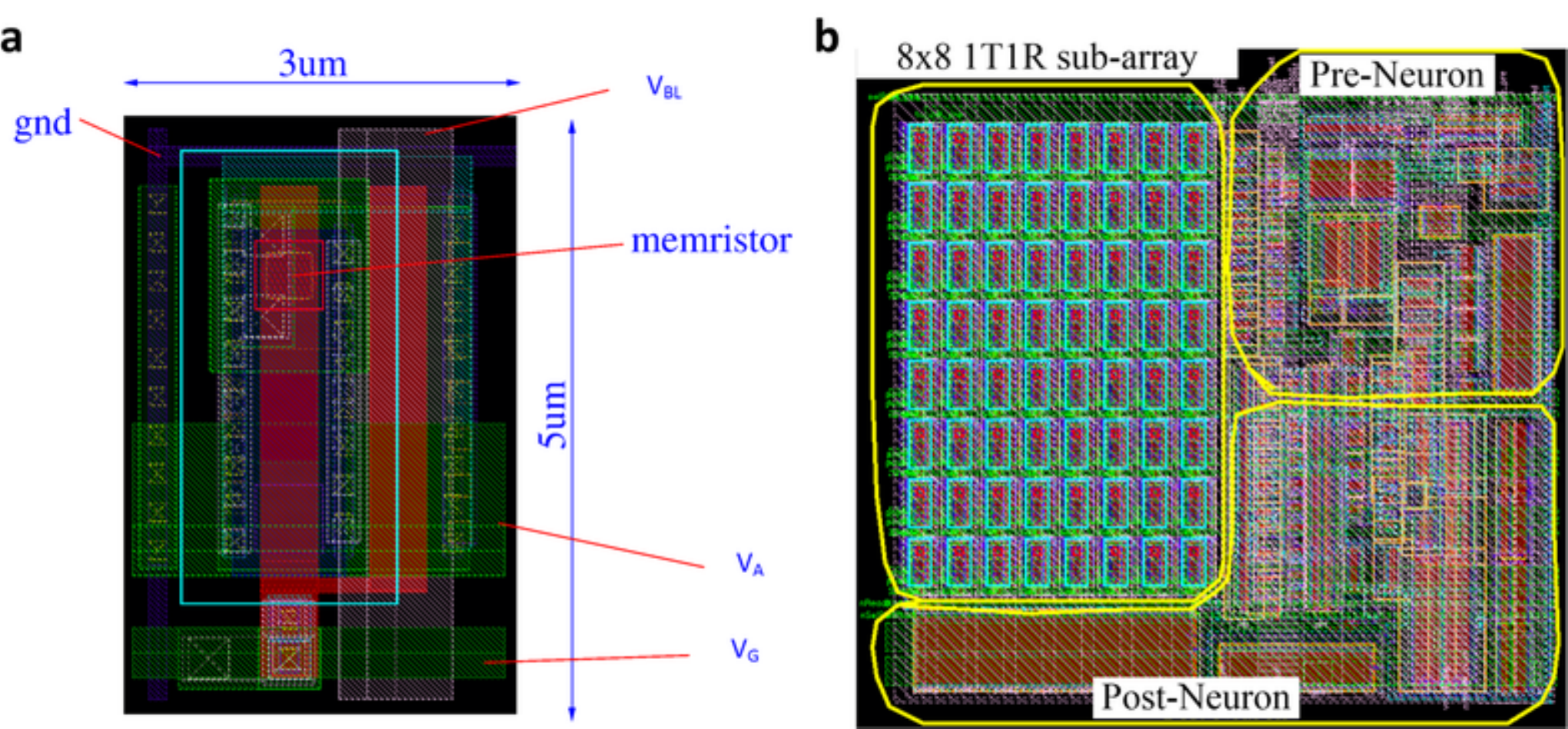}
\caption{a) 1T1R cell physical layout of size 3 $\mu m \times 5 \mu m$. Memristor area is about 200nm $\times$ 200nm. The thick-oxide nMOS transistor requires a width of W = 6.7 $\mu m$ and a length of L = 0.5 $\mu m$ and is split into two fingers. b) Layout of one macro-cell (or tile) with 8$\times$8 1T1R synapses, one pre-synaptic neuron and one post-synaptic neuron. The macro-cell has a size of 55 $\mu m \times 51 \mu m$, and can be assembled into arrays of macro-cells.}
\label{fig:Fig3}
\end{figure}

\begin{figure*}
\centering
\includegraphics[width=6in]{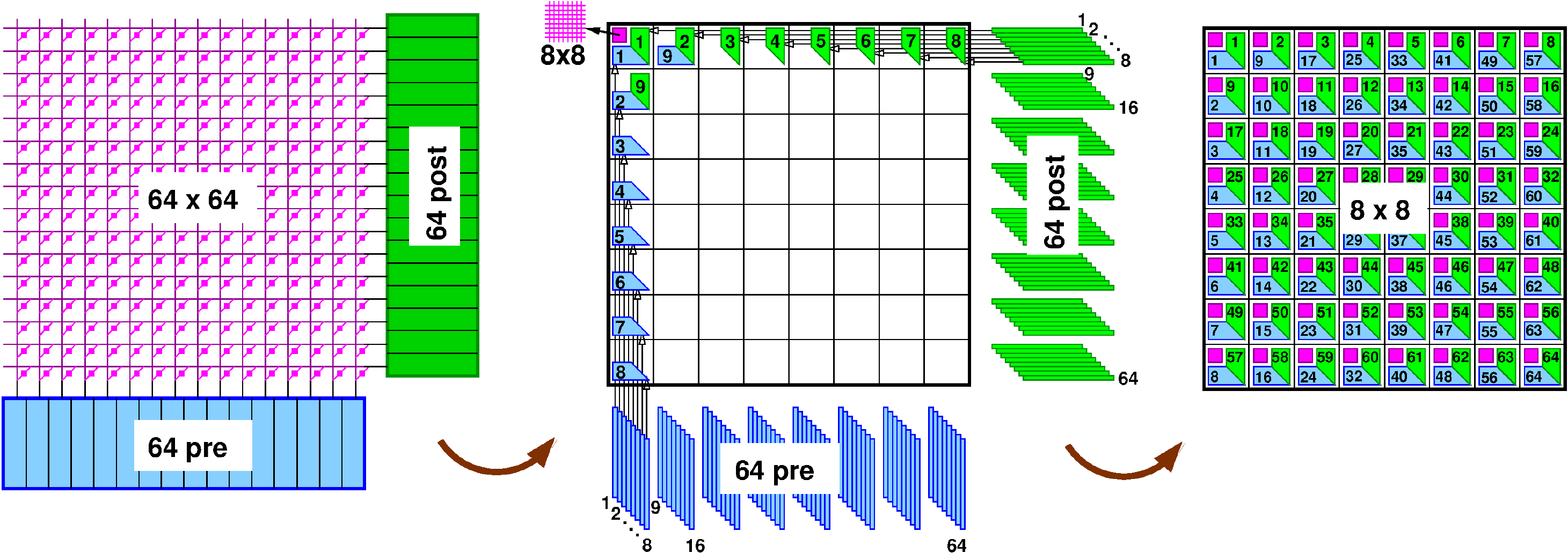}
\caption{Transformation of a 64$\times$64 memristive crossbar core into CMOL-like layout.}
\label{fig:Fig4}
\end{figure*}

In the chip presented in this paper, the memristor technology used was not yet ready to implement the CMOL concept exactly as described above, because each nano-scale memristor required one non-nanoscale NMOS transistor in series. This memristor-NMOS compound is called 1T1R. As a result, each synaptic element not only resulted in a non-nanoscale area of 3um $\times$ 5um (as illustrated in Fig. \ref{fig:Fig3}(a)), but also required CMOS real estate. In the present chip we therefore used a pseudo-CMOL approach, in an effort to adapt the original CMOL concept to the restrictions of the available technology. The transformation of the 64$\times$64 computational crossbar architecture into the corresponding physical pseudo-CMOL layout architecture is illustrated in Fig. \ref{fig:Fig4}. 

Fig. \ref{fig:Fig4} shows, on the left, a standard 64$\times$64 synaptic crossbar with 64 pre-synaptic neurons at the bottom and, on the right, 64 post-synaptic neurons. From Fig. \ref{fig:Fig3}(a) we know that in our technology each 1T1R synapse has a size and pitch of $3 \mu m\times 5 \mu m$. To maximize crossbar density, the 64$\times$64 synapse should therefore be placed at a 3$\mu m$ horizontal pitch and a 5$\mu m$ vertical pitch. However, this would require a presynaptic neuron layout with a 3$\mu m$ pitch, resulting in very thin but extremely long, practically non-viable, cells. Similarly, post-synaptic neurons would require a vertical pitch of 5 $\mu m$, resulting again in cells that would be extremely wide if viable at all\footnote{Given the layout rules of the technology used, the limited number of metal layers for routing (4), and the complex structures within the pre- and post-synaptic neurons in Fig.~\ref{fig:Fig3}(b), we are certain such a fine pitch layout would have been impossible, besides introducing massive parasitic couplings.}. To avoid this and produce a realistic, viable layout, we used the following pseudo-CMOL approach to draw the topological layout. The 64 pre-synaptic and post-synaptic cells were grouped together in groups of 8, as shown in the center of Fig. \ref{fig:Fig4}. The 64$\times$64 synapses were grouped into 8$\times$8 sub-groups, each having 8$\times$8 synapses. The layout was divided into 8$\times$8 identical macro-cells or tiles. Pre-synaptic neurons 1 to 8 were assigned to the first column of macro-cells. Pre-synaptic neurons 9 to 16 were assigned to the second column, and so on. Similarly, post-synaptic neurons 1 to 8 were assigned to the first row of macro-cells (see Fig. \ref{fig:Fig4}, center). Post-synaptic neurons 9 to 16 were assigned to the second row, and so on. The 64$\times$64 synaptic crossbar was drawn in such a way that the groups of 8$\times$8 synapses were drawn at maximum density of pitch 3$\mu m\times 5 \mu m$, but each 8$\times$8 group was drawn in the upper left corner of each macro-cell. This resulted in the layout shown in Fig. \ref{fig:Fig4} right, where each macro-cell includes an 8$\times$8 subset of the synaptic crossbar, one pre-synaptic neuron and one post-synaptic neuron. This way, the layout of the neurons could be made much more compact.
All synapses preserved their original 64$\times$64 synaptic crossbar connectivity, since the 64 horizontal rows and 64 vertical lines were unaltered, with pre-synaptic neuron 1 connecting only to the first column of synapses, pre-synaptic neuron 2 only to the 2nd column, and so on. Likewise, post-synaptic neuron 1 connected only to the first row of synapses, post-synaptic neuron 2 only to the second, and so on. The connectivity of synapses, pre-synaptic neurons, and post-synaptic neurons was thus preserved. The Appendix illustrates a more generic layout arrangement.

Fig. \ref{fig:Fig3}(b) shows the final layout of one such macro-cell, with the 8$\times$8 sub-array of 1T1R synapses at top left, the layout of the post-synaptic neuron at bottom right, and the pre-synaptic neuron at top right. The 8$\times$8 array of macro-cells, including 64$\times$64 = 4096 1T1R synapses + 64 pre-synaptic neurons + 64 post-synaptic neurons, occupies an area of $440 \mu m\times 408 \mu m$. 

\begin{figure}
\centering
\includegraphics[width=3.3in]{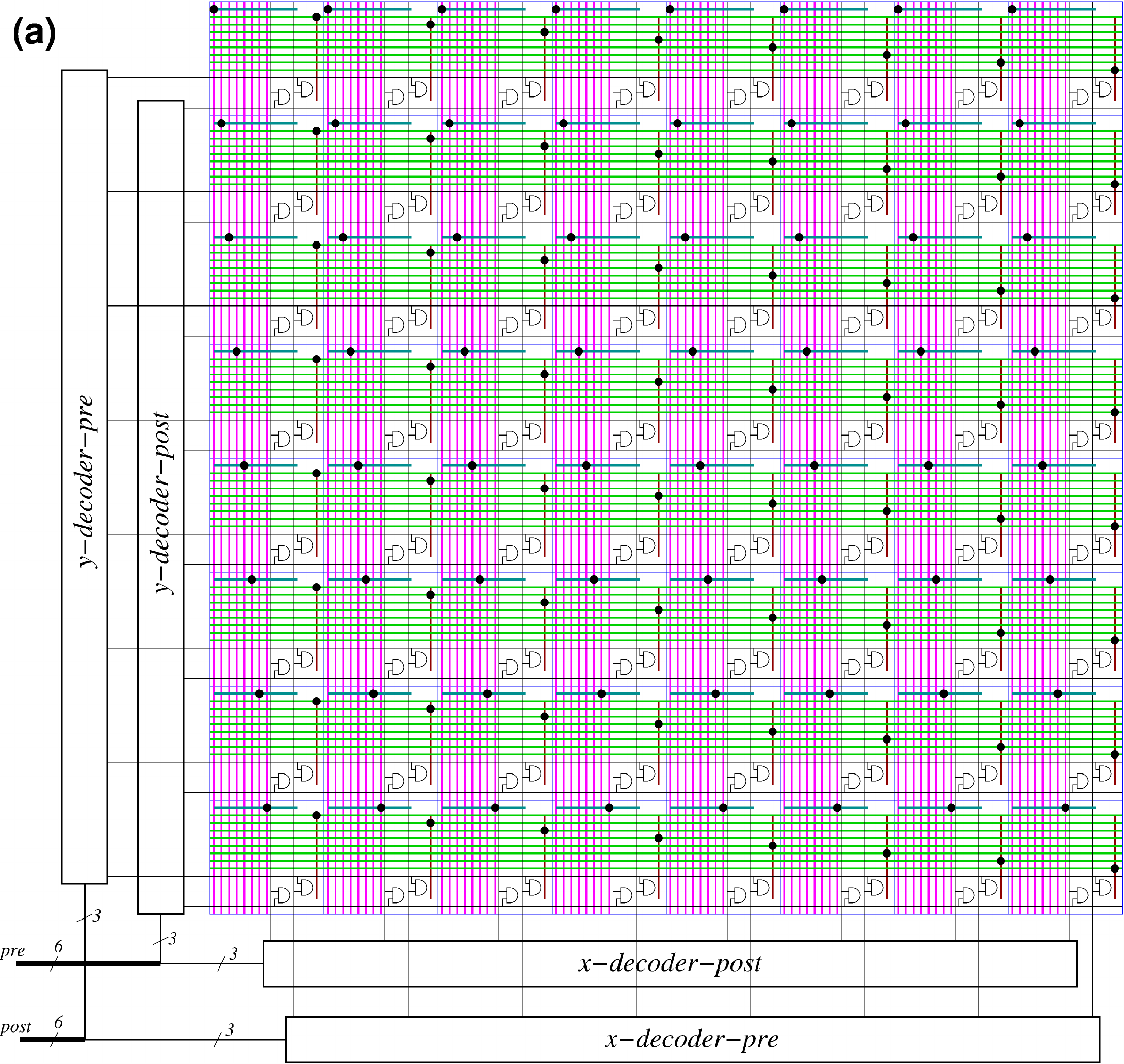} \includegraphics[width=2in]{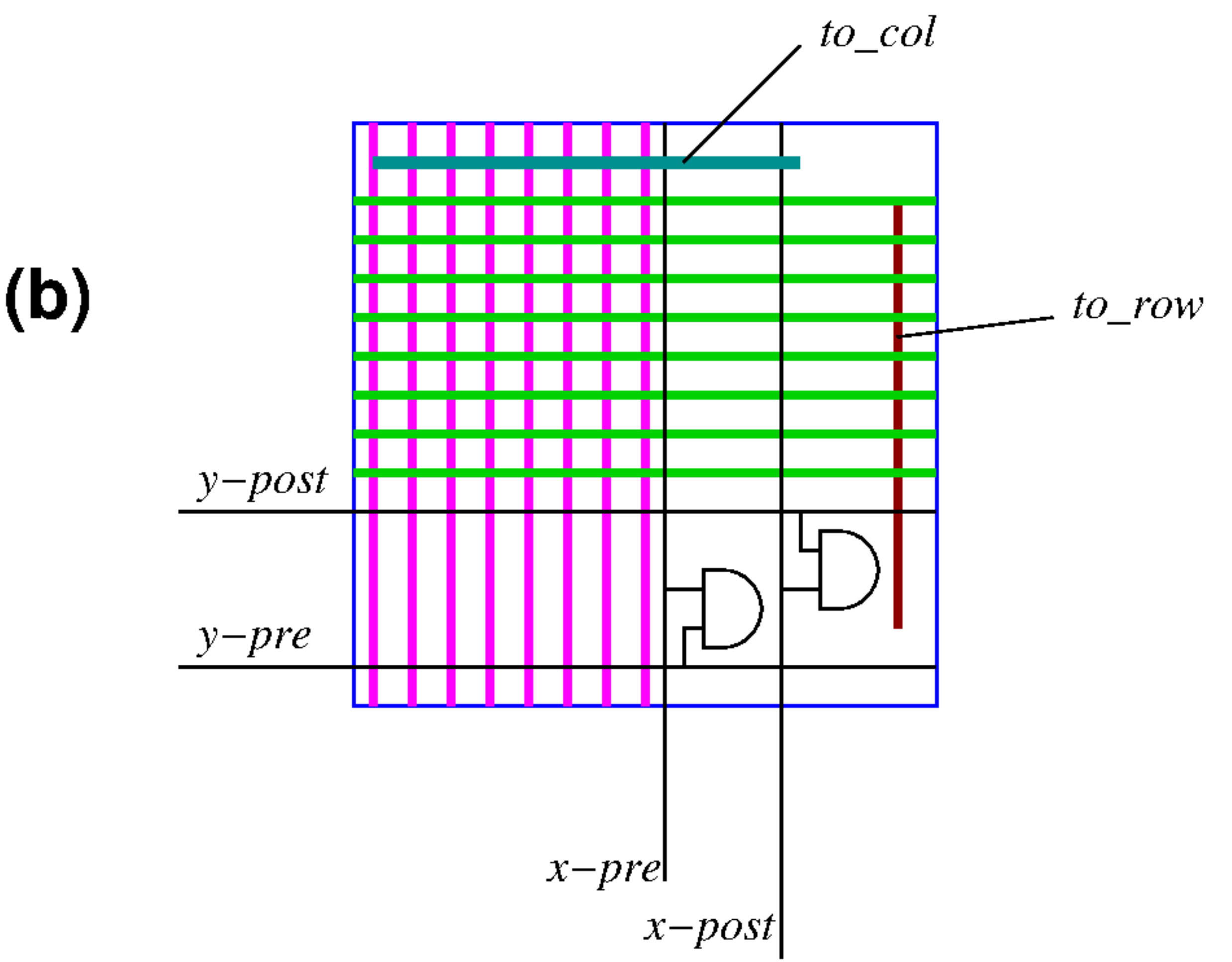} 
\caption{a) Simplified CMOL-core layout overview consisting of $8\times 8$ unit macro-cells. b) Layout overview of one macro-cell.}
\label{fig:Fig5}
\end{figure}
In practice, we do not use tilted lines to emulate CMOL. Fig.~\ref{fig:Fig5}(a) shows a simplified overview of the CMOL-core layout with $8\times 8$ unit macro-cells, and Fig.~\ref{fig:Fig5}(b) shows one such macro-cell. The vertical (pink) lines are fabricated in metal-3 while the horizontal (green) ones in metal-4. 1T1R devices are at their respective intersections. The unit macro-cell in Fig.~\ref{fig:Fig5}(b) includes a horizontal segment (fabricated in metal-2) named $to\_col$ that crosses all eight columns crossing the macro-cell, and a vertical segment $to\_row$ (fabricated in metal-1) that crosses all eight rows crossing the macro-cell. The layout of the macro-cell is unique and both segments $to\_col$ and $to\_row$ are not connected to any column or row. The macro-cells are connected at the next hierarchy level, as indicated with thick dots in Fig.~\ref{fig:Fig5}(a). This connectivity pattern complies with the pre- and post-synaptic neuron labeling presented in Fig.~\ref{fig:Fig4} right, and makes unnecessary to tilt the wires as in Fig.~\ref{fig:Fig1} (we tilt the ``lines'' of interconnecting dots instead). In our fabricated macro-cell, the pre-synaptic neuron output is hard-wired to segment $to\_col$, and the post-synaptic neuron output to segment $to\_row$. By adding a minimum of switches it is straight forward to swap these connections, and thus transpose the weight matrix~\cite{Merolla, Gert2020, Gert2021}. However, for the STDP application we were targeting this was not required. The peripheral decoders activate one pre-synaptic neuron in one of the macro-cells and one post-synaptic neuron in another one (or same) simultaneously, to access one specific memristor for forming, writing, erasing or reading. This would be the programming mode. In inference mode, one full column is activated at a time, while all post-synaptic neurons would be integrating.

Additionally, the event read-out scheme implemented is based on the query-driven approach \cite{query-driven}. In this approach post-synaptic neurons are accessed by querying their address, which allows to dynamically adjust their thresholds individually, a key feature for both (a) the implemented STDP rule discussed later and (b) allowing to calibrate for mismatch each neuron.

\section{Hardware implementation of Stochastic Binary STDP learning}

Spike-timing-dependent plasticity (STDP) \cite{Gerstner} is a biological mechanism for synaptic learning which refines the traditional Hebbian rule \cite{Hebb}. Different kinds of STDP algorithms have been proposed for machine learning and neuromorphic computing applications \cite{Delorme, Masquelier2009, Mozafari, Kheradpisheh}. After the emergence of memristor devices, it was demonstrated that STDP behavior emerges naturally when specially shaped pre- and post-synaptic spikes are applied at both sides of a memristor \cite{Linares2009}, opening up the possibility of implementing STDP learning in neuromorphic hardware \cite{Zamarreno, Prezioso2016}. 

\begin{figure}
\centering
\includegraphics[width=2.5in]{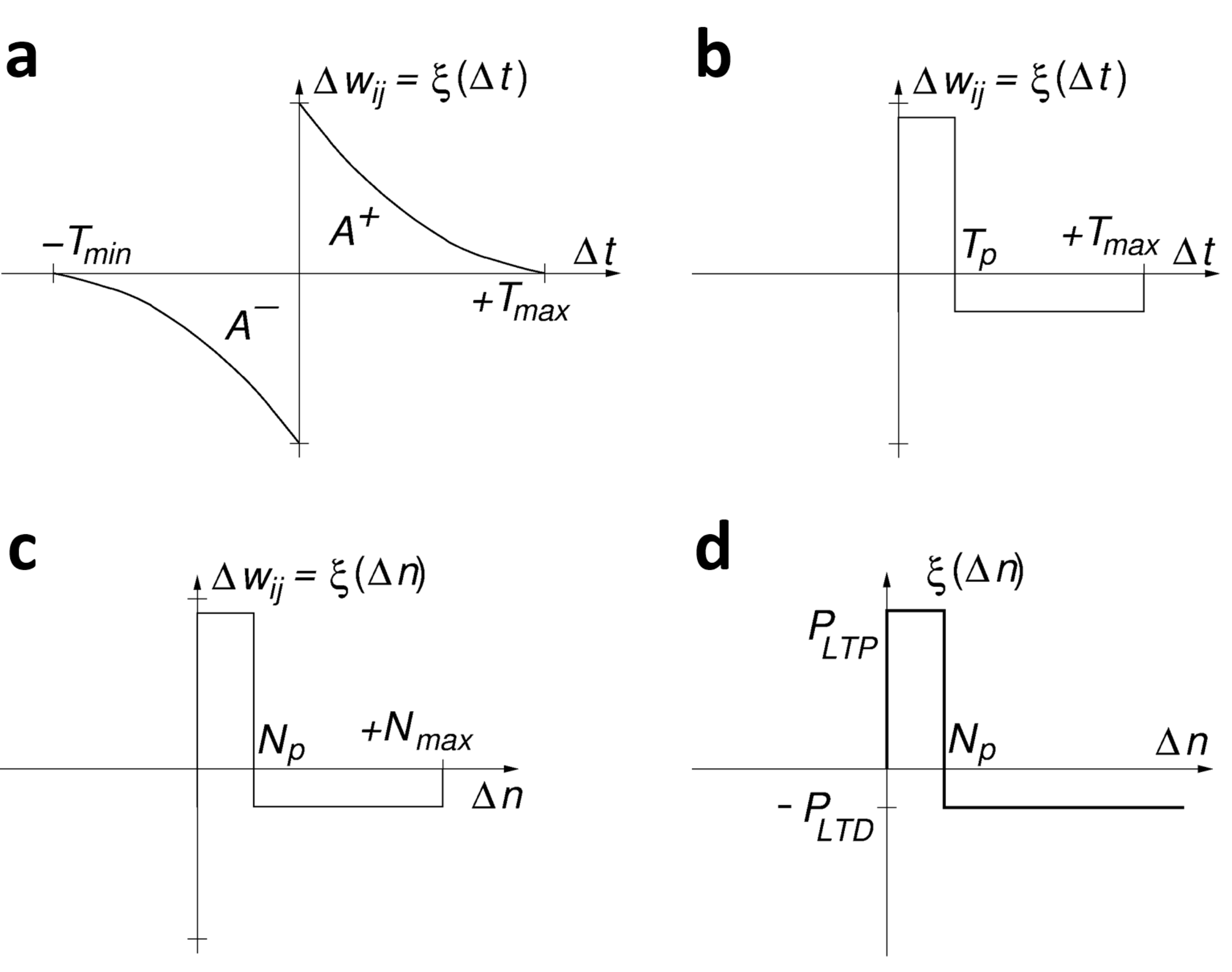}
\caption{Illustration of different SDTP update functions.}
\label{fig:Fig6}
\end{figure}

For one single synapse connecting two neurons, a generic STDP rule modifies the synaptic weight $w_{ij}$ as a function $\xi$ of the time difference $\Delta t$ between pre- and post-synaptic spikes ($\Delta t = t_{post} - t_{pre}$). The typical time-based STDP function is shown in Fig. \ref{fig:Fig6}(a), where positive (causal) $\Delta t$ increases the weight (synaptic potentiation) and negative (anti-causal) $\Delta t$ decreases the weight (synaptic depression). Smaller values of $\Delta t$ (in absolute value) produce larger weight changes. The right-hand side ($\Delta t > 0$) weight update is triggered whenever a post-synaptic neuron generates a spike, updating all synapses connected to it with a pre-synaptic neuron spike not older than $T_{max}$. Similarly, the left-hand side ($\Delta t < 0$) weight update is triggered whenever a pre-synaptic neuron generates a spike, updating all synapses connected to it with a post-synaptic neuron spike not older than $T_{min}$. Different computational simplifications for the update function $\xi$ have been proposed to boost computing speed. Some of these simplifications were exploited in this study for more efficient, compact hardware.

The first simplification, illustrated in Fig. \ref{fig:Fig6}(b), defines a narrow positive time window (0, $T_p$) where a fixed amount of potentiation is applied, while a fixed depression is applied otherwise, but only for positive values of $\Delta t$ between $T_p$ and $T_{max}$. Note that only the right-hand side is implemented, thus triggering weight updates only after post-synaptic spikes. An extension of this learning function proposes extending $T_{max} \rightarrow \infty$ \cite{Bichler, Suri, Querlioz}, forcing weight updates for all synapses connecting to a firing post-synaptic neuron, even in the absence of any prior pre-synaptic spike. 

In another simplification or abstraction, some researchers have proposed versions of the STDP function where the time variable has been removed and substituted by the rank-order of occurrence of spikes \cite{Bichler, Thorpe, Masquelier2007, Roclin}. This alternative consists of ordering the spikes as they are generated and ignoring their precise times. Fig. \ref{fig:Fig6}(c) shows the order-based learning function equivalent to Fig. \ref{fig:Fig6}(b), where $\xi$ is a function of the rank-order difference of occurrence of spikes n. This simplified updating rule still needs to be able to implement small changes $\Delta w_{ij}$ in the synaptic weights, and thus requires continuous analog or, at least, multi-valued memristive devices. In the study described in this paper, the memristive devices available were used as binary devices, meaning that a synapse was either fully ON or fully OFF. For this reason, we used the stochastic order-based STDP function shown in Fig. \ref{fig:Fig6}(d) \cite{CEA_SBSTDP, Yousefzadeh}. Every time a post-synaptic neuron generated a spike, this function assigned a probability of long-term potentiation $P_{LTP}$ to its synapses with a pre-synaptic spike separated by $0 < \Delta n < N_p$ in the rank-order, and assigned a probability of long-term depression $P_{LTD}$ to the rest of its synapses. The weights were updated only when a post-synaptic spike was generated, but they remained unchanged with the generation of pre-synaptic spikes \cite{Yousefzadeh}. As each synapse was implemented by a memristor, we were able to associate binary synaptic values to both the high resistive state (HRS) and the low resistive state (LRS) of the memristive device. 

\begin{figure}
\centering
\includegraphics[width=3.4in]{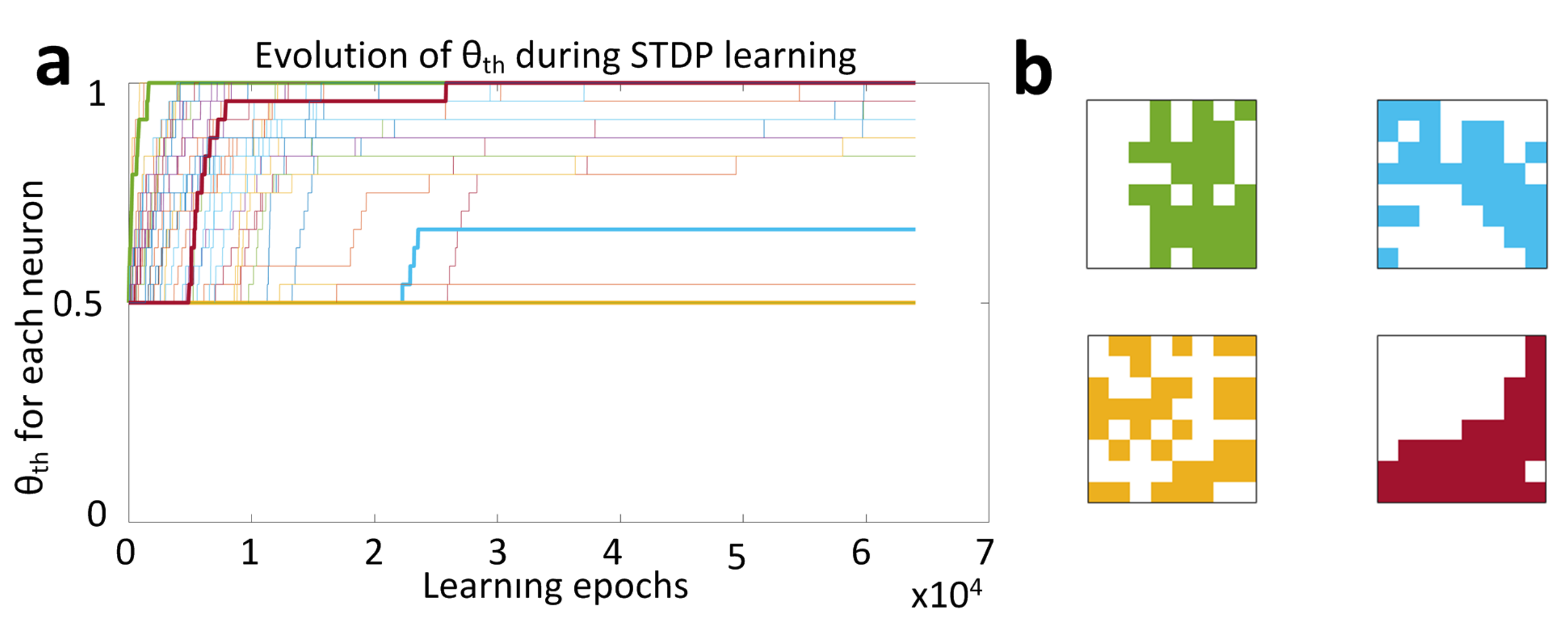}
\caption{a) Evolution of the neuron thresholds $\Theta_{th}$ of the 64 post-synaptic neurons during STDP learning using a behavioral model of the algorithm in Matlab. The input patterns for training were the ones shown in Fig. \ref{fig:Fig15}(a). A neuron state was defined between 0 (reset state) and 1. All neurons had an initial threshold $\Theta_{th} = 0.5$, and every time a single neuron generated a post-synaptic spike, the algorithm increased it by 0.04 (saturating at 1) while its input synapses were updated following the STDP rule. This plot illustrates the evolution of these thresholds. b) 2D representation of the final values of the synaptic weights associated to the 4 post-synaptic neurons in a) with the same colors. The white pixels represent memristors in HRS, while the colored ones correspond to LRS pixels. The yellow neuron did not learn any feature (random weights and minimum threshold), while the red neuron clearly learned a feature present in several letters. The green and blue neurons showed different intermediate learned features. }
\label{fig:Fig7}
\end{figure}

Although stochastic binary STDP was proposed originally in 2013 \cite{CEA_SBSTDP}, for realistic size systems a number of regularization techniques are required \cite{Yousefzadeh}. In the rest of this paper we will use acronym SB-STDP to refer to this regularized version.
To first illustrate SB-STDP learning we simulated it in Matlab using behavioral models, with 64 pre- and 64 post-synaptic neurons and 4,096 binary synapses with initial random values. Every time a post-synaptic neuron $j$ reached its threshold, a post-synaptic spike was produced, and the algorithm would check the list of its pre-synaptic spikes with $\Delta n < N_p$, identifying the pre-synaptic neurons $i_1$, $i_2$, etc. which had a direct influence on the spiking post-synaptic neuron. All synapses given by $(i_1, j)$, $(i_2, j)$, etc. would be activated with a probability $P_{LTP}$, while all other synapses connected to neuron $j$ would be deactivated with a probability $P_{LTD}$. At the same time, the threshold of output neuron $j$ would be potentiated to make that neuron less sensitive (e.g., it would become more difficult for this neuron to fire again). This way, after receiving enough pre-synaptic spikes, several post-synaptic neurons would become more specific to certain input features. Fig. \ref{fig:Fig7} illustrates MATLAB simulations of the proposed SB-STDP rule when the 64 input neurons received the simple binary 8$\times$8 visual input stimuli images shown later in Fig. \ref{fig:Fig15}(a). Fig. \ref{fig:Fig7}(a) shows the evolution of the neuron thresholds as their respective synapses were updated after firing. The neurons that experienced threshold increases were the ones learning patterns. In Fig. \ref{fig:Fig7}(a), we have highlighted 4 neurons with thicker lines. Three of them (the red, blue and green ones) underwent learning and the fourth (the yellow one) underwent no learning. The corresponding learned weights are shown in Fig. \ref{fig:Fig7}(b). As can be seen, the neurons that underwent learning resemble some of the input patterns from Fig. \ref{fig:Fig15}(a), while the one that did not undergo any learning remains in the initial random state. 

Adaptive threshold per individual neuron can be implemented in our core thanks to the query-driven read-out. This way, every time an output neuron is accessed for read-out, its dynamic threshold voltage is set individually during read-out (see around description of Fig.~\ref{fig:Fig9}(c) in Section IV).

\section{A Fully Integrated Monolithic Neuromorphic CMOS-Memristor CMOL-Core}

\begin{figure}
\centering
\includegraphics[width=3.4in]{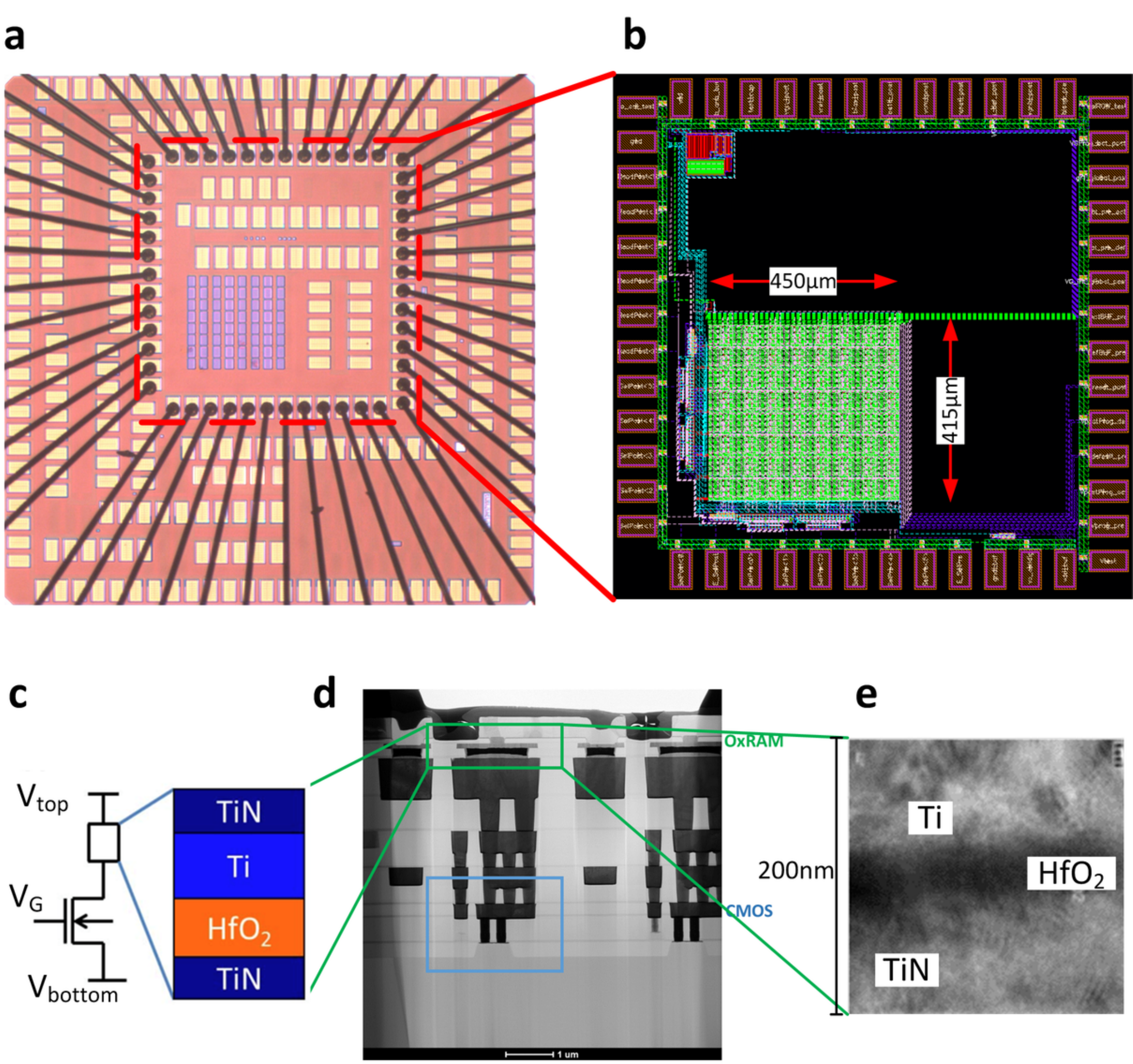}
\caption{Fully integrated hybrid CMOS-memristor chip.}
\label{fig:Fig8}
\end{figure}

In this work we designed, fabricated, and tested an integrated CMOL-like memristor-CMOS core with a memristor crossbar emulating synaptic connections and CMOS circuits implementing pre- and post-synaptic neurons. An array of 4,096 (64$\times$64) resistive memory devices in a one-transistor-one-memristor (1T1R) configuration was implemented, interleaved with the CMOS circuitry, in a CMOL-like geometrical layout approach, using a 130nm CMOS technology with 200nm Ti/HfOx/TiN memristors on top. Fig. \ref{fig:Fig8}(a) shows a photograph of the $5mm^2$ fabricated chip, which included several other test circuit structures, surrounded by a red dotted square, bonded on its package. The CMOL-core described in this paper is included in the central pad ring, the layout of which is shown in Fig. \ref{fig:Fig8}(b) (lower left green area). The other fabricated test structures are not shown in this layout drawing. The CMOL-core, with its 64 input neurons, 64 output neurons, and 4,096 1T1R synapses, occupied an area of $415\mu m\times 450 \mu m$. This area is slightly larger than the one given by multiplying the single macro-cell in Fig. \ref{fig:Fig3}(b) by 8$\times$8 ($408\mu m\times 440 \mu m$). This slight difference is given by small overhead introduced when building the final array.


Fig. \ref{fig:Fig8}(c) shows the 1T1R synaptic configuration required for the 130nm RRAM-CMOS technology used \cite{Valentian}.
The nMOS transistor could not be of minimum size as it needed to drive high currents to guarantee reliable erase operations. In this technology it also had to be a thick oxide (i.e., bulkier) transistor to allow for the higher voltages required for forming, writing, and erasing. 
A microscopic view of the monolithic integration of the 1T1R device within the CMOS substrate is shown in Fig. \ref{fig:Fig8}(d), together with a graphical description of the fabrication process used to build the memristive devices on top of the chip \cite{Valentian}. A microscopic view of the Ti-Hf$O_2$-TiN nanodevice is shown in Fig. \ref{fig:Fig8}(e).

\subsection{Description of on-chip operation}

\begin{figure}
\centering
\includegraphics[width=3.4in]{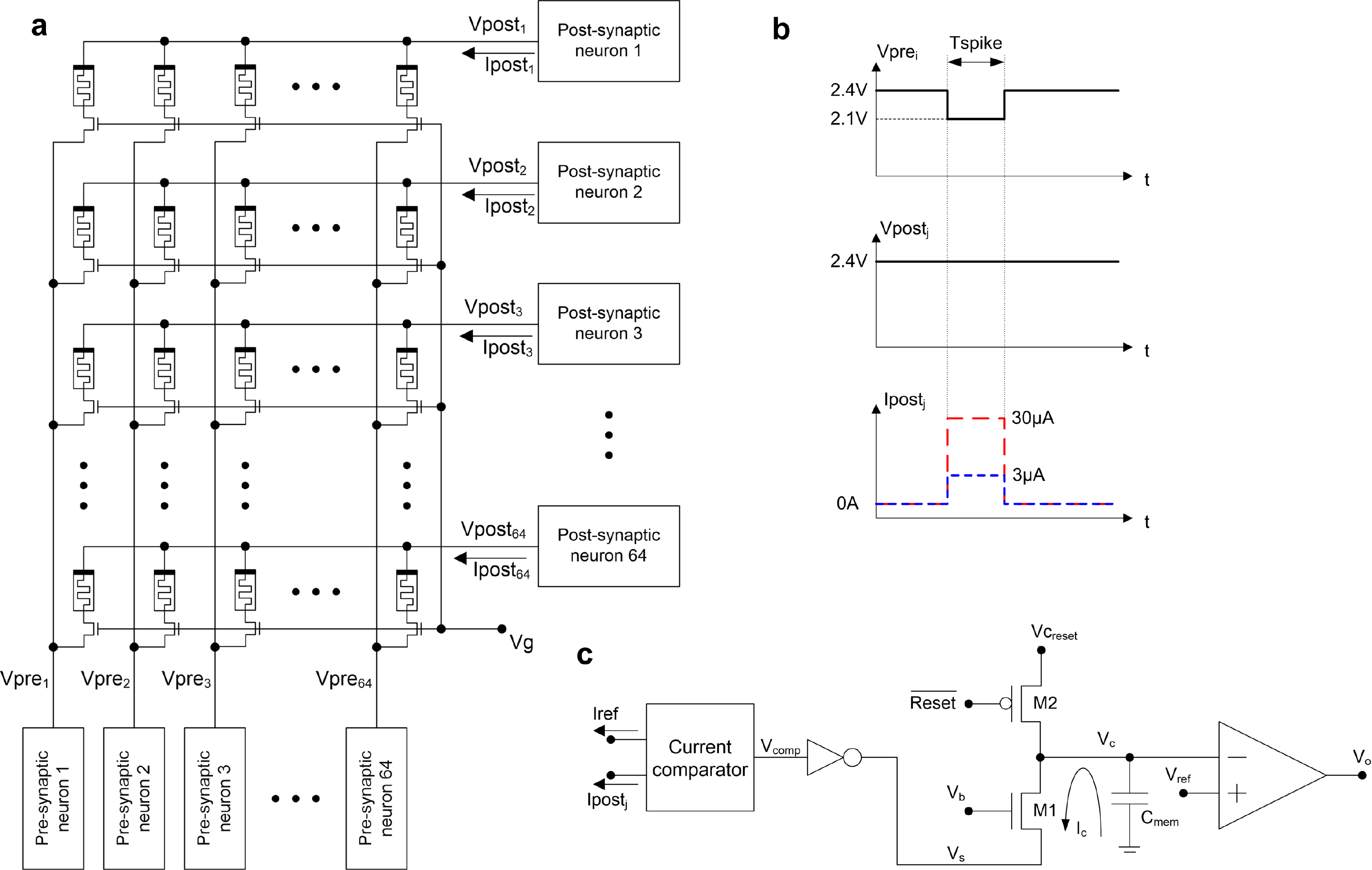}
\caption{Illustration of the operation of the fabricated on-chip memristor-CMOS CMOL-core.}
\label{fig:Fig9}
\end{figure}

Fig. \ref{fig:Fig9}(a) shows a simplified schematic diagram of the CMOL-core implemented on-chip, with 64 pre-synaptic neurons (each one connected to a vertical line $Vpre_i$) and 64 post-synaptic neurons (each one connected to a horizontal line $Vpost_j$), interconnected all-to-all through a crossbar of 64$\times$ 64 1T1R synaptic devices. The maximum current allowed through the memristors is controlled by Vg, which is shared by the gates of all the selector transistors. Every time a pre-synaptic neuron $i$ sends an input spike, it reduces the voltage of $Vpre_i$ from 2.4V to 2.1V for the duration of $T_{spike}$ as shown in Fig. \ref{fig:Fig9}(b), while $Vpost_j$ remains constant at 2.4V for all post-synaptic neurons j. This produces a voltage difference of 300mV in all the 1T1R devices in column i, generating a current $Ipost_j = 300mV/R_{ij}$, where $R_{ij}$ represents the resistance of memristor $ij$ in series with its selector. If the memristor is at HRS, and assuming an average memristor-selector series resistance of $100k \Omega$, the generated current will be $3 \mu A$ on average. If the memristor is at LRS, and assuming an average memristor-selector series resistance of $10k \Omega$, the generated current will be $30 \mu A$ on average, as illustrated in Fig. \ref{fig:Fig9}(b). A simplified diagram of the post-synaptic neuron is shown in Fig. \ref{fig:Fig9}(c), where the input current $Ipost_j$ is compared against a reference $Iref = 10 \mu A$. If the input current is larger than the reference, digital signal $V_{comp}$ will be high and digital signal $V_s$ will be low, enabling a certain amount of current $I_C$ through transistor M1 during $T_{spike}$, discharging capacitor $C_{mem}$ (125fF) by a given amount. Transistor M1 acts as a charge pump, with bias voltage $V_b$ setting the value of $I_c$, and thus setting the size of the charge packet $\delta q = I_c \times T_{spike}$ that discharges membrane capacitor $C_{mem}$. Each charge packet decrements the membrane capacitor voltage by $\delta V_c = \delta q/ C_{mem}$. Membrane capacitor voltage $V_c$ is compared against a reference voltage $V_{ref}$, output voltage $V_o$ indicating that the threshold has been reached. 
Neuron threshold voltage $V_{ref}$ can be set individually for each output neuron, thanks to the query-driven read-out, as required for the STDP rule described in Section III, as illustrated in Fig.~\ref{fig:Fig7}(a).\footnote{The query-driven read-out consumes $2\times n_{lev}$ clock cycles per output neuron, where $n_{lev}$ is the number of threshold levels per neuron. In our case, we used an FPGA clocked at 50MHz, so that each clock cycle is 20ns. For a common threshold for all neurons ($n_{lev}$=1), the query-driven read-out for the full core (64 neurons) takes $2.56\mu s$. We typically set $n_{lev}=13$ levels, as in Fig.~\ref{fig:Fig7}(a), resulting in a worst case $33.3\mu s$.} Consequently, the query-driven read-out, if implemented per core, is negligible compared to the typical learning dynamics of spiking neural networks.
Transistor M2 is used to reset the capacitor voltage to $Vc_{reset} > V_{ref}$. Note that in the physical neuron circuit, incoming spikes discharge the membrane capacitance and the voltage is reset to a maximum value after spike generation. This is the opposite of what happens in biology or in standard computational models, although this is transparent from the system level operation point of view. When making the neuron less sensitive by “potentiating” its threshold, the physical threshold voltage $V_{ref}$ should therefore be progressively decreased, as opposed to what happens with the computational threshold illustrated in Fig. \ref{fig:Fig7}(a). 

\section{Experimental Results}

\subsection{Test setup}

\begin{figure}
\centering
\includegraphics[width=2.5in]{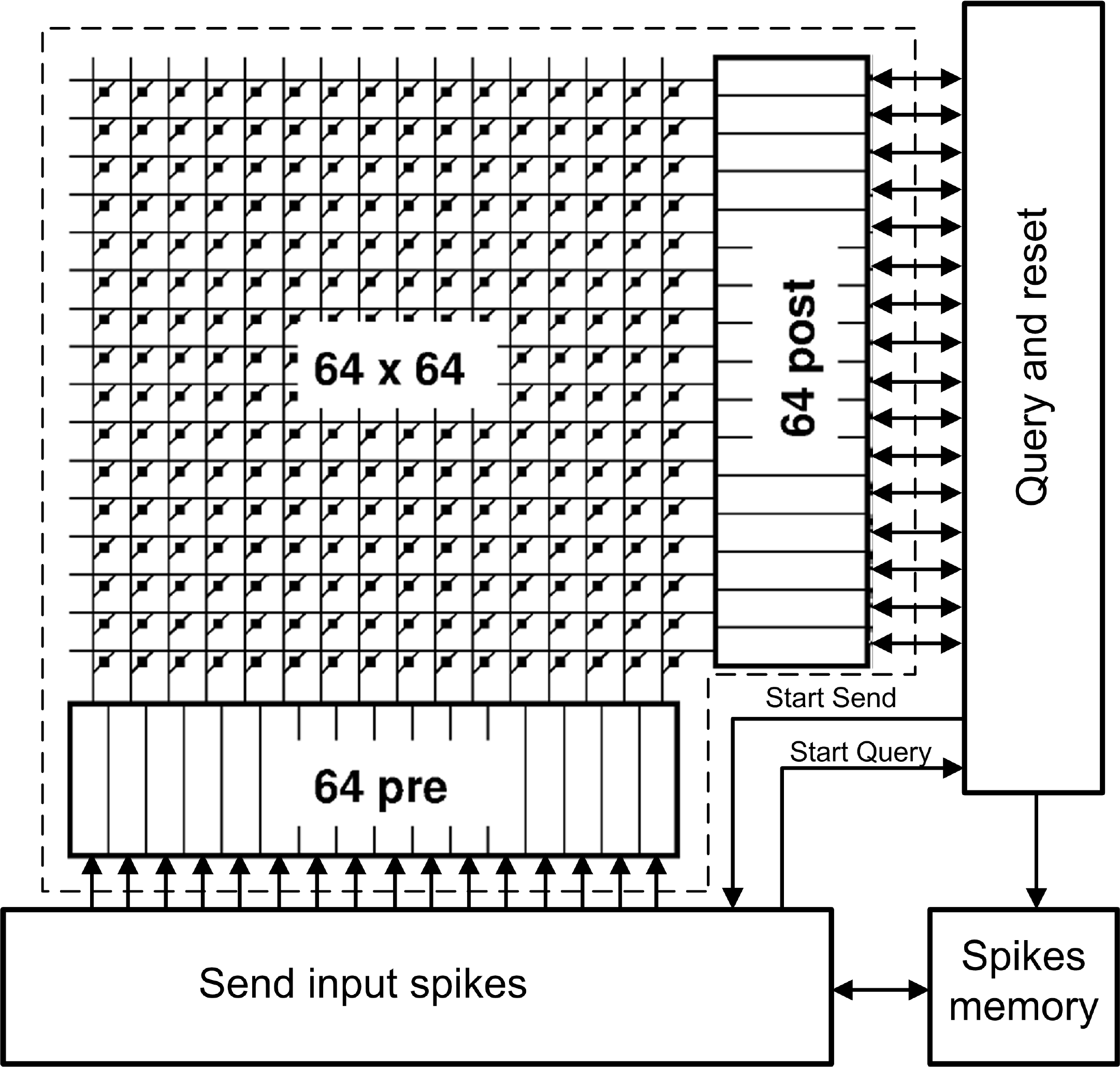}
\caption{Illustration of the operation of the fabricated on-chip memristor-CMOS CMOL-core.}
\label{fig:Fig10}
\end{figure}

Fig. \ref{fig:Fig10} shows a diagram of the complete hardware platform used to perform the tests described in this work. The dashed area includes the blocks implemented on-chip while the rest of the blocks were implemented on an FPGA connected to the chip. As mentioned earlier, the fabricated chip CMOL-core included 64 pre- and 64 post-synaptic neurons with a crossbar of 4,096 1T1R memristors with all-to-all connectivity. On the FPGA, the following blocks were included:

\begin{itemize}
    \item A block which sent input spikes to the pre-synaptic neurons on the chip.
    \item A block which queried all post-synaptic neurons on the chip in order to identify which ones reached their threshold, and reset them.
    \item A memory block where the pre-synaptic stimulation input spikes were written before the experiment and where the post-synaptic spikes were saved during the experiment for further analysis. 
\end{itemize}

This experimental setup was used to characterize the memristive network for two different scenarios: a first set of experiments where the memristors were programmed off-line beforehand for inference operations and a second experiment where we demonstrated on-line SB-STDP learning performance for unsupervised feature learning and consequent classification \cite{Yousefzadeh}.

\begin{figure}
\centering
\includegraphics[width=2.5in]{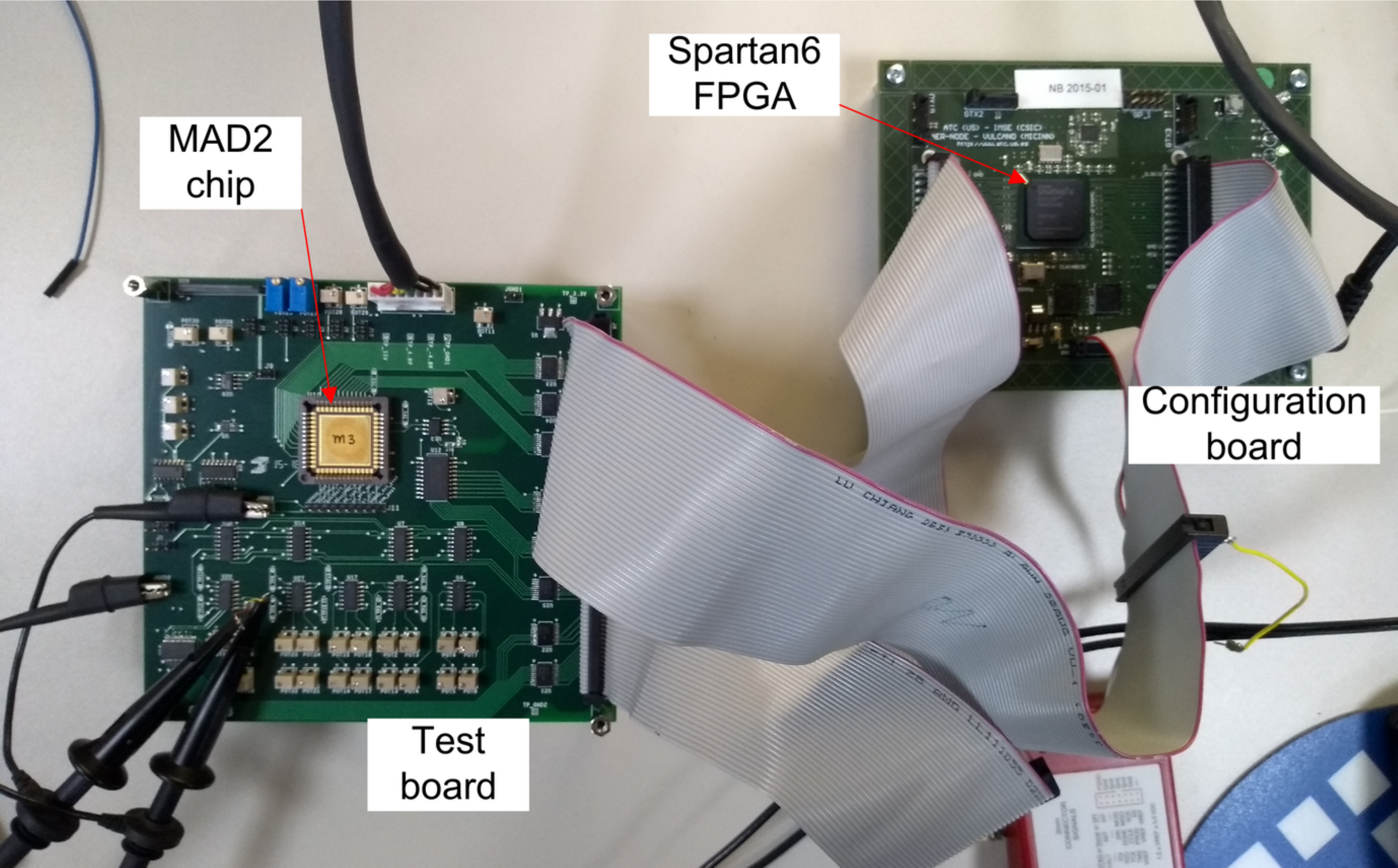}
\caption{Illustration of the operation of the fabricated on-chip memristor-CMOS CMOL-core.}
\label{fig:Fig11}
\end{figure}

A photograph of the test setup is shown in Fig. \ref{fig:Fig11}, with the two different boards: a test PCB including the fabricated chip and some external components used to generate the necessary voltage and current biases, and a configuration PCB with a Spartan 6 FPGA. This FPGA was used to run different algorithms and configure the chip for each specific test. The tests involved modifying the state of each individual memristor and implementing inference and stochastic binary STDP algorithms. 
Operations associated with modifying or reading the state of the memristors were controlled by the FPGA by enabling and disabling certain switches in the test PCB in order to apply specific voltages to the 1T1R devices for a given time. There were 4 different operations: Form, Write, Erase and Read. Each operation required the application of a voltage pulse to one individual 1T1R structure. This pulse was defined by its duration ($T_{spike}$) and the 3 voltages $V_{top}$, $V_{bottom}$ and $V_g$ (see Fig.~\ref{fig:Fig8}(c)). Table \ref{table:operations} details the pulses applied to Form, Write, Erase and Read the memristors. 

\begin{table}[h!]
\centering
\caption{Description of memristor operations}
\label{table:operations}
\begin{tabular}{ |c|c|c|c|c|c| } 
 \hline
  & Form & Write & Erase & Read & Inference \\ [1ex]
 \hline
 Duration & $10 \mu s$ & $100 ns$ & $100 ns$ & $10 \mu s$ & $200 ns$ \\ 
 Vtop (Vpost) & $4.8V$ & $2.4V$ & $0V$ & $2.4V$ & $2.4V$\\ 
 Vbottom (Vpre) & $0V$ & $0V$ & $4.8V$ & $2.1V$ & $2.1V$ \\ 
 Vg & $1.5V$ & $1.5V$ & $4.8V$ & $3.5V$ & $3.5V$ \\ [1ex]
 \hline
\end{tabular} 
\end{table}

\subsection{Characterization of memristors}

\begin{figure}
\centering
\includegraphics[width=3.4in]{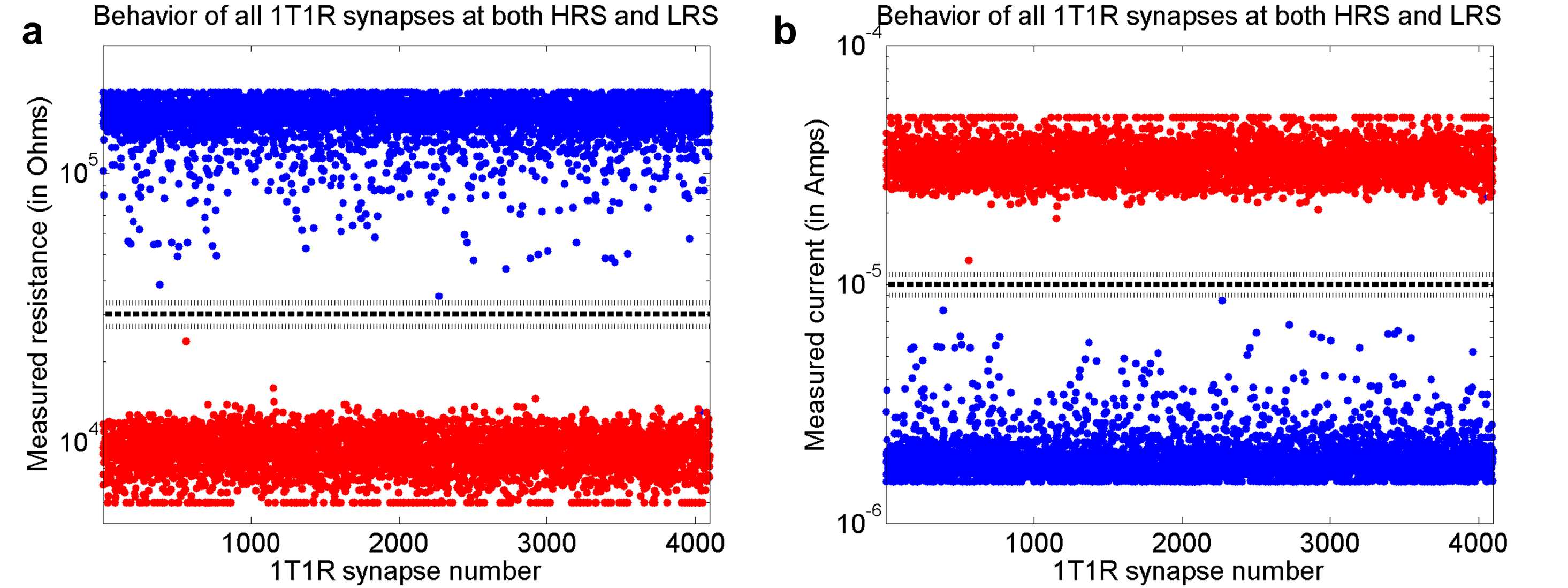}
\caption{Characterization of all 1T1R synapses at both HRS and LRS. }
\label{fig:Fig12}
\end{figure}

Once all 4096 memristor devices had been formed, ‘Erase’ pulses were applied sequentially to all of them repeatedly. The resistance of each memristor was then measured, obtaining the values represented by the blue dots in Fig. \ref{fig:Fig12}(a), corresponding to HRS. ‘Write’ pulses were then also applied sequentially to all the devices repeatedly, and the resistances were measured again, obtaining the values represented by the red dots in Fig. \ref{fig:Fig12}(a), corresponding to LRS. Most HRS values were above $100k \Omega$ and most LRS values were below $15k \Omega$. The current sensing circuit used measured resistance values of between $6k \Omega$ and $200k \Omega$. Values outside this range produced saturation. The horizontal line in Fig. \ref{fig:Fig12}(a) represents a $30k \Omega$ resistance (with $±10\%$ mismatch-induced range), which was considered as a limit between HRS and LRS. Plot Fig. \ref{fig:Fig12}(b) represents the physically measured currents driven by each 1T1R synapse to the post-synaptic neuron when the corresponding pre-synaptic neuron sent an input spike. The red dots represent the current driven when the memristors were in LRS, the blue dots represent the current driven when they were in HRS. A current comparator in the post-synaptic neuron (see Fig.~\ref{fig:Fig10}(c)) was used to check if the received current was larger than a threshold, discriminating between an LRS memristor (active synapse) and an HRS memristor (inactive synapse) to obtain binary synaptic behavior. 

This way, each neuron integrates a fixed charge packet when the memristor is in LRS, and does not integrate any charge packet when in HRS. Therefore, the post-synaptic neuron computing precision is totally insensitive to memristor variability, as long as the neuron can safely discriminate between LRS and HRS. The down-side of this approach is that only one column can be activated at a time during inference. However, in Spiking Neural Networks (SNNs) input events typically come in event-by-event. Thus, if the column can be activated for a short time (200ns in our case, as explained next in C.1) this is not a severe drawback.

As shown in Fig. \ref{fig:Fig12}(b), a threshold of $10 \mu A$ (with $\pm 10\%$ mismatch-induced deviation) was sufficient to distinguish clearly between the two cases. In this figure, the full 4k memristor array was set repeatedly to HRS or LRS until the measured resistances were clearly separated into two non-overlapping HRS/LRS regions.

\subsection{Template matching inference experiments}

Since each post-synaptic output neuron was connected to all 64 pre-synaptic input neurons through 64 1T1R MOS-memristor synapses, for the inference experiments it was possible to program a template with 64 pixel values for each output neuron, one value for each synaptic connection. In this experiment each input neuron encodes the value of a pixel in an 8$\times$8 binary pattern. Since we had 64 output neurons, a total number of 64 templates could be programmed on the memristive crossbar, so that every time the input neurons provided one of the 64-bit input patterns, a single output neuron would match the most similar pattern and fire an output spike. Each inference experiment therefore required a set of 64 8$\times$8 binary images. Results are given below for two template matching inference experiments, each one with a different choice of input patterns: the first had human-meaningful patterns of varying numbers of active pixels, and the second had human-meaningless patterns but with a fixed number of active pixels.

\subsubsection{Experiment using a set of selected ASCII characters}

\begin{figure*}
\centering
\includegraphics[width=5in]{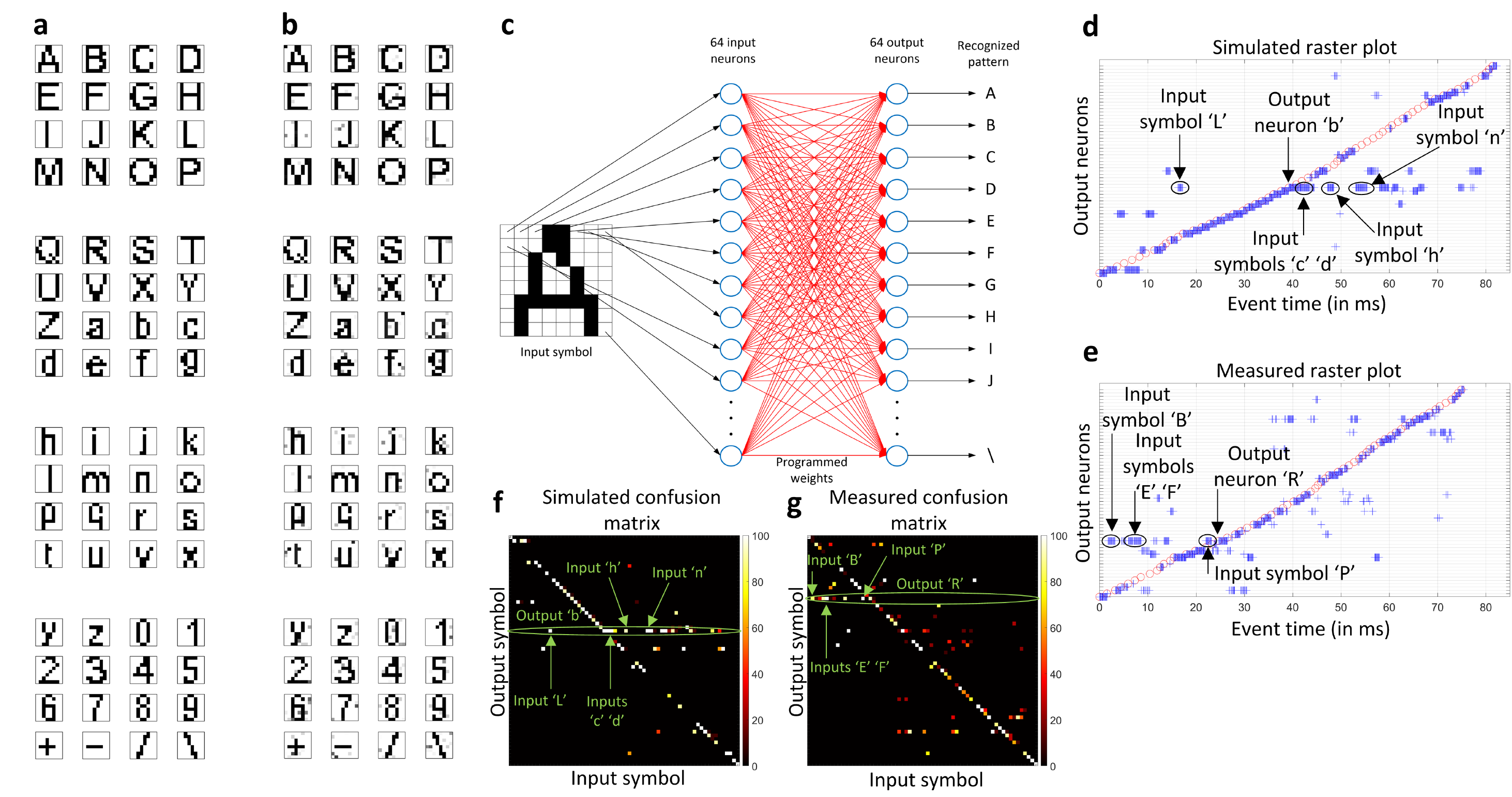}
\caption{Experimental demonstration of an inference example on the CMOS-memristor chip. 
}
\label{fig:Fig13}
\end{figure*}

In a first experiment, a set of 64 ASCII characters was considered, including 25 capital letters, 25 lower case letters, 10 digits (‘0’ to ‘9’), and plus (‘+’), minus (‘-‘), slash (‘/’) and backslash (‘$\backslash$’) characters. For all of them, the corresponding Arial font character was down-sampled to obtain an 8$\times$8 pixel image with 1-bit color depth (binary image). The 64 characters obtained are shown in Fig. \ref{fig:Fig13}(a). 
Having different numbers of active pixels, and with a high level of overlap between templates, identifying 64 different characters is not a trivial problem even under ideal conditions. For this reason, a perfect performance cannot be expected for this experiment. However, the aim of this test is to compare the experimental results with those obtained from behavioral simulations.

Each of these 8$\times$8 characters was encoded in 64 memristors connected to a single output neuron, so that each memristor encoded one pixel and each output neuron was programmed to recognize one of the 64 characters. For that purpose, the FPGA was programmed to read all 64$\times$64 pixel values sequentially, and ‘Write’ or ‘Erase’ each memristor, depending on whether the corresponding pixel was ‘1’ (black) or ‘0’ (white). Once all 64$\times$64 memristors in the chip had been programmed, their resistances were measured, obtaining the templates shown in Fig. \ref{fig:Fig13}(b). Ideally, all memristors associated to a black pixel should have had a resistance value of around $10k \Omega$ (LRS), while all those associated to a white pixel should have had a resistance value of around $100k \Omega$ (HRS). In practice, we observed a certain degree of stochasticity in the behavior of the devices: sometimes they were not programmed properly or attained a resistance value somewhere between LRS and HRS. In any case, all programmed characters were noisy but clearly recognizable by visual inspection. 

A schematic description of this experiment is shown in Fig. \ref{fig:Fig13}(c). The input image was converted into a spike train which was sent to the input layer of neurons (each pixel connected to an input neuron). Each neuron in the output layer corresponded to a certain template (a character, in this example). The all-to-all synaptic connections corresponded to the memristive crossbar, and were programmed with the pattern values. 

After all the synaptic values had been programmed in the memristive crossbar, the inference experiment consisted of sequentially presenting all 64 input characters to the CMOL-core. To do this, each character was converted into a spike train (one 200ns spike per black pixel) which was sent to the corresponding pre-synaptic input neuron by the FPGA. The spike list for each character was repeated several times. 
This experiment was first simulated in Matlab using a behavioral model of the system.
For each input spike, the current flowing through all 64 memristors connected to the corresponding input neuron was calculated using each spikes’ physically measured resistance. This way, 64 input currents were obtained for all 64 output neurons. For each output neuron, the input current was compared to a threshold $I_{th}$ ($10 \mu A$). When  the current was larger than this threshold, the capacitor voltage $V_c$ representing the membrane potential, was updated by $\delta V_c$. If the new $V_c$ value reached the neuron threshold $V_{th}$, an output spike was sent and all neurons were reset.

This experiment was simulated introducing a random variation to $\delta V_c$ in order to emulate the mismatch between neurons in the chip. The raster plot obtained is shown in Fig. \ref{fig:Fig13}(d), where the Y-axis represents the 64 output neurons (indicating each one’s programmed character) and the X-axis represents spike times in milliseconds. Each red circle indicates the beginning of the playback of each input character, and each blue cross represents an output spike. 
This realistic simulation showed a number of incorrect spikes which did not match the input character being processed. Symbol ‘b’, for example, had several black pixels in common with other symbols, and in the simulation corresponding to this plot, the output neuron associated to symbol ‘b’ had a larger $\delta V_c$ value. This neuron was therefore likely to attract other similar input symbols, as represented in Fig. \ref{fig:Fig13}(d) by input symbols ‘L’, ‘c’, ‘d’, ‘h’ or ‘n’, among others. The corresponding confusion matrix is shown in Fig. \ref{fig:Fig13}(f).

The confusion matrix represents the complete output activity generated by the network, with the 64 input characters on the X-axis and the 64 output neurons on the Y-axis. For each input character, a number of output spikes were generated, in general by different output neurons. Each position in a given column encoded the ratio of events generated by each output neuron while a given input character was being applied, with white representing $100\%$ of the events and black $0\%$. For example, and focusing again on output neuron ‘b’, there were many active pixels in the corresponding row in Fig. \ref{fig:Fig13}(f), with several input symbols wrongly recognized as ‘b’. The ratio of correct spikes over the total number of output spikes obtained was $60.78\%$ in this simulation. After performing 100 simulations with different values of $\delta V_c$ (random mismatch), a statistical distribution of this ratio of correct output spikes was obtained, with a mean value of $52.93\%$ and a standard deviation of $7.09\%$.

The same experiment was then repeated in hardware. For this, the spike list was written on the FPGA, which was programmed to the spikes to the chip while scanning and resetting the output neurons. The generated output spikes were saved on the FPGA, obtaining the raster plot shown in Fig. \ref{fig:Fig13}(e), which shows very similar behavior to the one observed in simulation. The corresponding confusion matrix is shown in Fig. \ref{fig:Fig13}(g). When we computed the performance of the template matching experiment, a ratio of $57.91\%$ correct spikes was obtained, only slightly lower than the simulated result. Here too there were some incorrect classifications. For example, the ‘R’ symbol had many pixels in common with other symbols like ‘B’, ‘E’, ‘F’ or ‘P’. That explains why all those input symbols (and others) appear misclassified as ‘R’ in Fig. \ref{fig:Fig13}(e) and (g).

\subsubsection{Experiment using random irregular connected shapes}

\begin{figure*}
\centering
\includegraphics[width=5in]{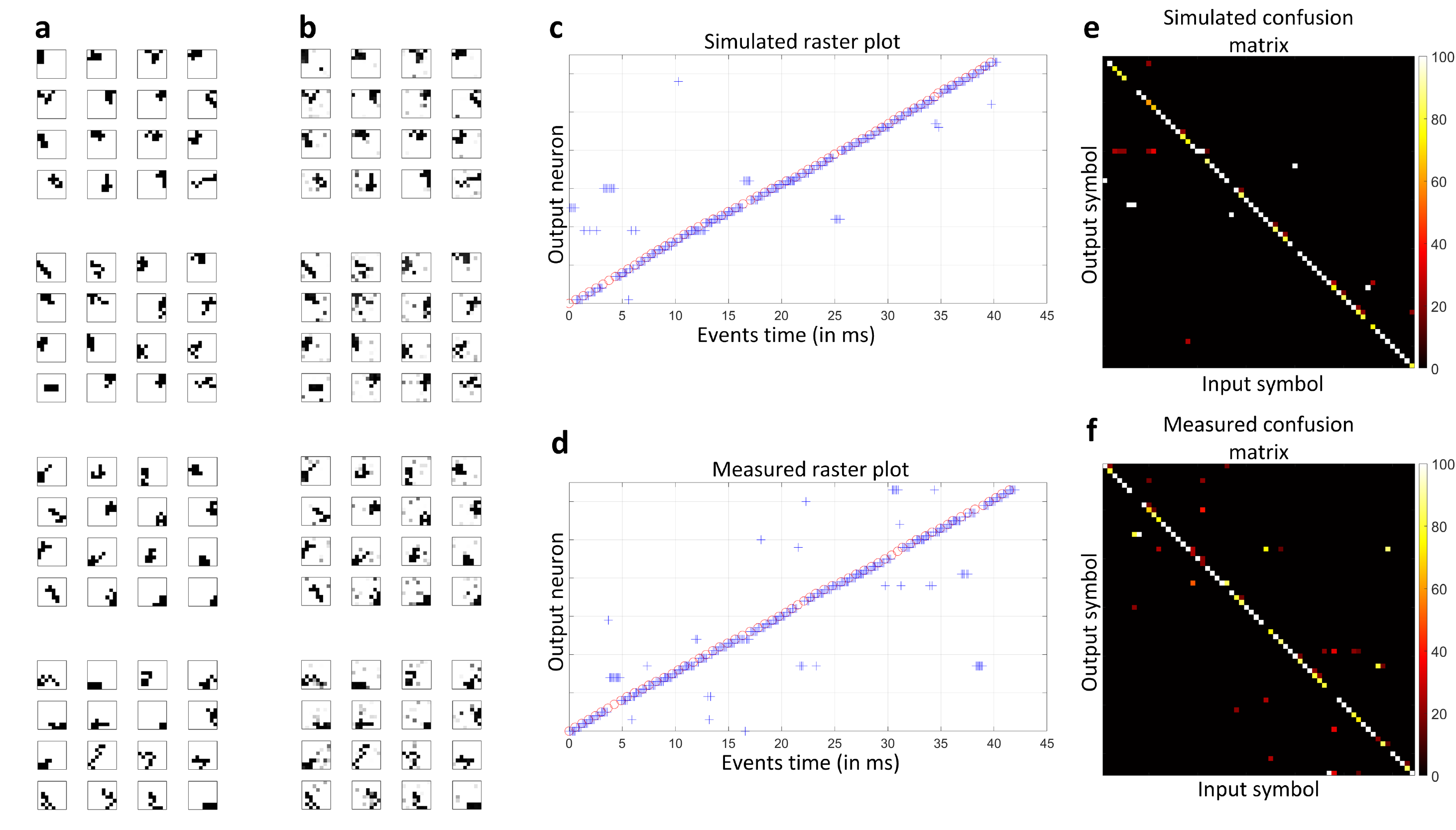}
\caption{Experimental demonstration of the pre-programmed inference example on the CMOS-memristor chip, with random connected input patterns of equal length. 
}
\label{fig:Fig14}
\end{figure*}

For a second inference experiment, 64 random irregular shapes were considered. 64 input images were created with 8$\times$8 pixels, each irregular shape being formed by only 8 pixels. To provide a reasonable separation between all 64 images, each shape was generated starting from a different seed pixel and letting it grow randomly inside the image, maintaining continuity. Fig. \ref{fig:Fig14}(a) shows the chosen set of 64 images. 

The 64 irregular shapes were used as inputs for this second template matching experiment. First, they were encoded in the memristive crossbar by programming LRS if a template pixel was black or HRS if it was white. The resistive values of all the memristors were measured, obtaining the results shown in Fig. \ref{fig:Fig14}(b).

After all the synaptic values had been programmed in the memristive crossbar, the inference experiment consisted of sequentially presenting all 64 input symbols to the network, as in the previous experiment. 

This experiment was first simulated using the same Matlab model, obtaining the raster plot shown in Fig. \ref{fig:Fig14}(c) and the confusion matrix shown in Fig. \ref{fig:Fig14}(e). The performance of this simulation was computed by calculating the ratio of correct spikes over the total number of output spikes, obtaining $82.73\%$.

This experiment was also repeated on-chip, obtaining the raster plot shown in Fig. \ref{fig:Fig14}(d) and the confusion matrix shown in Fig. \ref{fig:Fig14}(f). Behavior was very similar to that observed in simulation. When the performance of the template matching experiment was computed, a ratio of $81.79\%$ correct events was obtained, slightly lower than the simulated result.

This second experiment showed better performance than the first, both in simulation and experimentally, because the input patterns were generated with a fixed number of active pixels and a low ratio of active versus total pixels, while maximizing the separation between patterns.

\subsection{Total Chip-Core Power and Inference Energy Measurements}
By measuring the chip’s power supply current during repetitive inferences, we obtained an average total current consumption for the CMOL-core of $I_{Vdd} = 2.3mA$. Power supply voltage for this chip was Vdd=4.8V.
Therefore, the total power chip consumption at maximum inference rate was 11mW.
Inferences had a periodicity of $T_p = 220ns$ (200ns at high level and 20ns at inter-spike stand-by level). Each input spike acted simultaneously on $n_{syn} = 64$ memristors. Consequently, the overall chip-total effective energy per synaptic operation is given by

\begin{equation}
  E_{SOP} = I_{Vdd} \times V_{dd} \times T_p / n_{syn} = 37.95pJ
\end{equation}

For comparison, Table \ref{table:comparison} shows the energy per synaptic operation for other related works reported in the literature. In addition, we also list the equivalent effective total charge drained from power supply per synaptic operation $\Delta Q_{SOP} = E_{SOP}/V_{DD}$, for which our approach outperforms the rest.

\begin{table*}[h!]
\centering
\caption{Comparison with some reported state-of-the-art spiking neural processing chips}
\label{table:comparison}
\begin{tabular}{ |c|c|c|c|c|c|c| } 
 \hline
  & TrueNorth \cite{Merolla} & Loihi \cite{Davies} & LETI \cite{Valentian} & DYNAPS \cite{Moradi} & ODIN \cite{Frenkel} & \textbf{This work} \\ [1ex]
 \hline
 Technology & Digital & Digital & Mixed-signal & Mixed-signal & Digital & Mixed-signal \\ 
 Feature size & 28nm & 14nm & 130nm & 180nm & 28nm & 130nm \\ 
 Power supply & 0.7V & 0.75V & 1.2V & 1.3V & 0.55V & 4.8V \\ 
 Weight storage & SRAM & SRAM & RRAM & SRAM & SRAM & RRAM \\ 
 $E_{SOP}$ & 27pJ & 105pJ & 180pJ & 30pJ & 13pJ & 38pJ \\ 
 $\Delta Q_{SOP}$ & 38.6pC & 140pC & 150pC & 23.1pC & 23.6pC & 7.92pC \\ [1ex]
 \hline
\end{tabular} 
\end{table*}

\subsection{Stochastic-Binary STDP learning experiment for classification}

A more complex classification experiment was used to demonstrate the SB-STDP feature learning algorithm on the CMOL-core, following previous computational work~\cite{Yousefzadeh}. A set of 4 binary letters (A, B, C, and D) with a resolution of 32$\times$32 pixels was used as input data. Each 1024-pixel input letter was divided into 16 sub-images, each with a resolution of 8$\times$8 pixels, as shown in Fig. \ref{fig:Fig15}(a). Each 8$\times$8 sub-image, which represented a “feature” of the full letter, could be translated directly into spike trains connected to the 64 input neurons on the chip, as illustrated in Fig. \ref{fig:Fig15}(b). Each input letter was therefore transformed into a sequence of 16 “features” represented by spike trains. The structure of the neural network is represented in Fig. \ref{fig:Fig15}(b), with two layers of 64 neurons fully connected and implemented on-chip (SB-STDP layer) followed by a classification layer implemented off-chip with 4 output classification neurons corresponding to input letters A, B, C, and D.
For the SB-STDP experiments, we used two different phases: a first phase to train the SB-STDP feature extraction layer, and a second phase to train the classifier layer. We also trained the classification layer with untrained random weights in the SB-STDP feature extraction layer. This way, the classifier layer was trained under two different conditions for the prior feature-extraction layer: one random and one trained through SB-STDP. This would allow us to assess whether SB-STDP was improving the discrimination capability of the classifier: that is to say, whether SB-STDP could learn features that were sufficiently relevant to help the classifier layer achieve better performance. We will now briefly describe the learning phases of the feature extraction layer and the classifier layer.

\subsubsection{Stochastic-Binary STDP feature-extraction layer learning phase}

The SB-STDP learning algorithm consists of several steps. Initially, random weights are assigned to all synapses in the SB-STDP layer (for each output neuron, half of the weights are set randomly to ‘0’ and the other half to ‘1’). In the next step, the network processes a list of input spikes so that the algorithm can search for correlations between pre- and post-synaptic spikes and modify the weights, allowing the output neurons to learn certain features. While processing the input spikes, the algorithm keeps an ordered list of the last pre-synaptic neurons which have sent a spike. The last 64 pre-synaptic spikes are therefore saved, and all the neurons included in that list are kept as the list of firing input neurons. Every time a post-synaptic neuron reaches its threshold, the algorithm uses that list to identify all synapses with correlating pre- and post-synaptic activity for that output neuron. All correlated synapses connected to that neuron are thus set to ON (LRS programmed) with a certain probability $P_{LTP}$, and the uncorrelated synapses connected to that neuron are set to OFF (HRS programmed) with a certain probability $P_{LTD}$. The threshold of the active output neuron is then potentiated in order to make that neuron more selective, making it more difficult for it to learn new patterns. Finally, the algorithm reads the total number of LRS memristors connected to the output neuron and modifies the state of some of them in order to keep that number equal to a given fixed value $N_{LRS}$. If the measured value is lower than $N_{LRS}$, some HRS memristors are randomly chosen to be set to ON to compensate the difference. If the measured value is larger than $N_{LRS}$, some LRS memristors are randomly chosen to be set to OFF to compensate the difference. These steps were followed to implement the SB-STDP learning algorithm \cite{Yousefzadeh} for the first feature extraction layer.

\subsubsection{Classification layer learning phase}

To validate the online SB-STDP learning algorithm, we implemented the network shown in Fig. \ref{fig:Fig15}(b) with the input patterns presented in Fig. \ref{fig:Fig15}(a). A single experiment consisted of sequentially processing the 16 8$\times$8 fractions of letter A, followed by the 16 fractions of letter B, and then those of letters C and D. Considering each 8$\times$8-pixel fraction as an input stimulus, the whole experiment involved sequentially processing 64 stimuli, each of which belonged to either letter A, B, C or D. After training both the SB-STDP and the classification layer, the aim of the experiment was for each classification neuron A, B, C and D to generate output spikes while the corresponding input letter was being processed, to indicate that the letter had been recognized. If classification neuron A generated a spike while the network was processing a fraction of letter A, that spike was considered correct, otherwise incorrect. 

The weights in the SB-STDP layer were trained directly on the chip first, following the algorithm described above, while the weights in the classification layer were trained using the following rule. Each classification neuron$j$ ($j$ = A, B, C, D) received input spikes through 64 synapses with analog weights $w_{ij}$ ($i$ = 1 … 64). Once the SB-STDP layer had been trained,
the 4 input letters were presented with the SB-STDP algorithm disabled.
Once the 16 stimuli associated to letter A had been presented, each SB-STDP layer output neuron $i$ ($i$ = 1 … 64) in the chip generated a total number of spikes $N_{iA}$. When all the stimuli associated to letters B, C and D had been presented, the numbers of spikes $N_{iB}$, $N_{iC}$ and $N_{iD}$ were obtained for each of the 64 post-synaptic neurons in the SB-STDP layer. If all those spikes were projected to the classification neurons, we were able to add all numbers of spikes generated by the SB-STDP layer for each input letter, obtaining $N_A$, $N_B$, $N_C$ and $N_D$ ($N_j = \Sigma N_{ij}$). Using these numbers, we assigned each synaptic weight the value $w_{ij} = N_{ij}/N_j$, where $i$ = 1… 64 represents the index of an SB-STDP layer output neuron and $j$ = A, B, C, D represents the index of a classification layer output neuron. 

After using the spikes generated by the chip to train the classification layer, we used the weights trained for the classification layer to process those spikes again off-chip, obtaining a list of spikes generated by the 4 classification neurons. For each input letter presentation A, B, C and D, we counted the number of spikes generated by each classification neuron, obtaining $N_{classA}$, $N_{classB}$, $N_{classC}$ and $N_{classD}$. Considering that symbol $j$ would be recognized if $N_{classj}$ was larger than any of the others, a decision was made for each input symbol based on the largest number of spikes among all the $N_{classj}$. This decision could be correct or incorrect, depending on whether $j$ matched the input symbol or not. The decisions obtained were $D_A$, $D_B$, $D_C$ and $D_D$ ($D_j$ = 1 when correct and $D_j$ = 0 when incorrect). These numbers gave us a number of correct spikes $N_{correct}$ (spikes generated by the correct neuron) and a number of incorrect spikes $N_{incorrect}$ (spikes generated by the other 3 incorrect neurons together). After processing the 4 input letters, we obtained the ratio of correct spikes as $R_{ev} = N_{correct}/(N_{correct} + N_{incorrect})$, and the ratio of recognized letters as $RR = \Sigma_j D_j/4$. 

This second phase for training weights in the classification layer was repeated for random untrained weights in the SB-STDP layer. For that purpose, all input letters A, B, C and D were processed by the SB-STDP layer with learning disabled, and the spikes obtained were used to train and validate the classification layer, following the same procedure as that described above and obtaining network performance measurements ($R_{ev}$ and RR) which can be compared with respect to those obtained using SB-STDP learning. This allowed us to assess the benefits of using SB-STDP feature learning for the first layer in helping the classifier layer.

To validate the network implemented on-chip, we first simulated its behavior with Matlab and then compared the experimental results with the simulation results.

\subsubsection{Simulation results}

First, we implemented the complete SB-STDP learning algorithms described above, including the training of both the SB-STDP and classification layers, in Matlab. In the simulations, we included a simple model of the mismatch in neuron current $I_c$ that we had found between post-synaptic neurons on-chip. This mismatch was the dominant non-ideal effect, as it produced a different discharging slope in the capacitor voltage for each neuron. Due to this effect, several neurons needed to integrate different numbers of input spikes before reaching the threshold. We modeled this mismatch in Matlab by adding a random number to the capacitor current. More specifically, we used a membrane capacitor current with a mean value of $I_c = 10nA$ and a standard deviation of $2.5nA$, representing a mismatch of $25\%$. This value produced a simulated behavior consistent with the experimental data. To compensate the mismatch, we modified the threshold of each output neuron when SB-STDP learning was disabled. 

In this framework, one single simulation consisted of the following steps:

\begin{itemize}
    \item Step 1. We programmed the memristor crossbar with random values, assigning each memristor either HRS or LRS with equal probability. 
    \item Step 2. With these random values, we first processed all the input spikes with SB-STDP learning disabled, obtaining a list of output spikes generated by the 64 post-synaptic neurons. Using these spikes, we trained the classification layer only. Once the classification layer was trained, we again presented all the input patterns to the full network. From the spikes generated by the classification neurons, we computed the ratio of correct spikes $R_{ev}$ and the recognition rate RR, both with random weights in the SB-STDP layer.
    \item Step 3. We then reprocessed all the input spikes with SB-STDP learning enabled, obtaining a new distribution of memristor values in the crossbar. 
    \item Step 4. With these learned values, we repeated step 2, processing the input spikes once again with SB-STDP learning disabled and using the obtained spikes to train the classification layer. Finally, the spikes were processed with the classification neurons. From these final spikes, we computed $R_{ev}$ and RR with the learned weights in the SB-STDP layer. 
    \item Step 5. We evaluated the SB-STDP learning algorithm by comparing $R_{ev}$ and RR before and after learning.
\end{itemize}

\begin{figure*}
\centering
\includegraphics[width=5in]{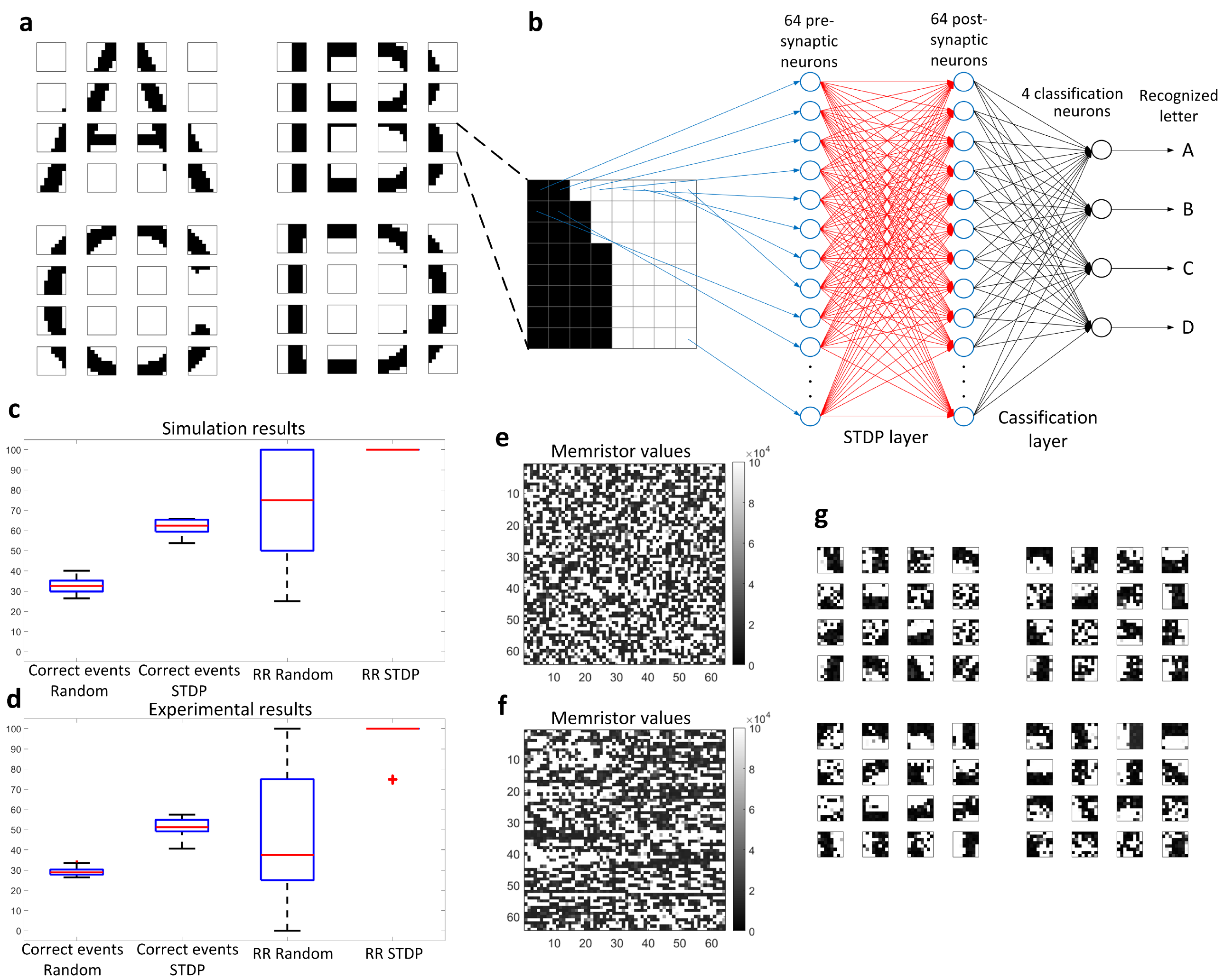}
\caption{Experimental demonstration of online STDP learning for classification with the CMOS-memristor CMOL-core. 
}
\label{fig:Fig15}
\end{figure*}

We repeated these steps 10 times, starting with different initial random weight distributions, to obtain a statistical characterization of the learning method. The results obtained are shown in Fig. \ref{fig:Fig15}(c), where we can see the statistics of the 10 experiments. Fig. \ref{fig:Fig15}(c) also compares the ratios of correct spikes $R_{ev}$ and recognition rate RR in the system with random weights (Step 2) with those in the system with SB-STDP learning enabled (Steps 3-4). Each plot shows the median value in red, the range between 25 and 75 percentile in a blue box, and the maximum and minimum values with black whiskers. As can be seen, when we used the network with random values in the memristor crossbar, we obtained a median value of correct spikes of around $35\%$, with a median recognition rate RR of below $80\%$. When we enabled the SB-STDP learning algorithm, however, we obtained a considerable improvement, with median values of more than $60\%$ for correct spikes and $100\%$ for RR. Our next objective was to reproduce this behavior with the fabricated chip CMOL-core. 

\subsubsection{Experimental results}

Here we repeated the same tests using the hardware setup shown in Figs.~\ref{fig:Fig10}-\ref{fig:Fig11}. We followed the 5 steps described in the previous section 10 times, each time starting with a different initial random weight distribution in the memristor crossbar. The results obtained are shown in Fig. \ref{fig:Fig15}(d), where we can see the statistics of the 10 experiments. Fig. \ref{fig:Fig15}(d) also compares the ratios of correct spikes $R_{ev}$ and recognition rate RR in the network with random weights with those in the network with SB-STDP learned weights. Each plot shows the median value in red, the range between 25 and 75 percentile in a blue box, and the maximum and minimum values with black whiskers. The red crosses represent outliers. As can be seen, when we used the network with random values in the memristor crossbar, we obtained a median value of correct spikes of around $30\%$ and a median recognition rate RR of below $40\%$. When we apply the SB-STDP learning algorithm, however, we obtained a considerable improvement, with median values of over $50\%$ for $R_{ev}$ and, more importantly, $100\%$ for the recognition rate RR. These results show that the neuromorphic CMOL-core can learn online visual patters in its memristor crossbar by applying an unsupervised SB-STDP learning algorithm to learn more relevant features, achieving very good final classification performance in pattern recognition tasks.

The behavior of the SB-STDP algorithm can be analyzed in more detail by observing the weight values before and after learning. Fig. \ref{fig:Fig15}(e) shows the measured resistance of all the memristors in the CMOL-core after programming random initial weights. The main difference with the simulation plots is that here we see no exact binary values, because the figure shows the precise resistance value measured for all the memristors. Although ideally we expected a binary behavior for these devices, in practice we found that both Writing and Erasing operations did not always produce an ideal effect. For this reason, we did not measure binary resistance values for all the memristors programmed at LRS or HRS, but instead observed a wider distribution, including around $3\%$ of the 1T1R devices in the border region between the two binary values. We also found that occasionally some memristors did not adopt the required state, i. e., they did not write or erase when the corresponding pulse was applied. These non-idealities might be caused by internal defects in some devices, by the aging effect of modifying their state too many times, or by a combination of both. More importantly, however, our work demonstrated the robustness of the network architecture and the learning algorithm, which provided very good recognition results even with memristors subject to such limitations. 

Fig. \ref{fig:Fig15}(f) shows the resistance values of all the memristors after SB-STDP learning. Note the difference with the initial random distribution in Fig. \ref{fig:Fig15}(e). Fig. \ref{fig:Fig15}(g) shows the learned patterns for each output neuron arranged as 8$\times$8 images. It can be seen how most of the patterns learned by the output neurons correspond to some of the 8$\times$8 fractions of letters presented in Fig. \ref{fig:Fig15}(a), as also indicated by the simulation results. Each neuron therefore learned a certain pattern corresponding to one of the 4 input letters, allowing the classifier to combine all 64 neurons properly and project them into the 4 classification neurons. This is a major improvement on the recognition rate shown in Fig. \ref{fig:Fig15}(d).

\section{Conclusions and Future Outlook}

This paper presents a full monolithic CMOS-memristor CMOL-like realization of a neuromorphic event-driven spiking neural network chip computing CMOL-core. The CMOL-core comprises 64 pre-synaptic neurons, 64 post-synaptic neurons and 4,096 1T1R memristive synapses. The monolithic technology used is CEA-LETI 130nm fabrication process, which uses ST-Microelectronic’s 130nm CMOS up to the metal-4 layer with added 200nm HfOx memristors on top plus an additional metal-5 interconnect layer. The CMOL-core and its 4k memristive synapses have been exhaustively tested and experimental computing applications have been demonstrated, including template matching experiments and regularized stochastic binary spike-timing-dependent plasticity (STDP). All the transistors used in the chip design were thick oxide transistors powered at 4.8V. This was a conservative decision made at the beginning of the design process, as it was the first time the technology had been made available outside LETI. The experimentally measured chip-total energy efficiency gave a figure of merit of 38pJ per synaptic operation (or, equivalently, its inverse as 26.3G synaptic operations per second per Watt). This energy efficiency can, in principle, be easily reduced by a factor of 3-to-4 by using lower voltage transistors, powered at 1.2V, for all circuit operations except for forming, writing, and erasing memristors. In this technology, the footprint of one 1T1R synapse is $3 \mu m \times 5 \mu m = 15 \mu m^2$. LETI is currently finalizing plans to use these memristors in 28nm \cite{Chang} and 22nm FDSOI technologies \cite{Hraziia}. The 1T1R footprints for these technologies are expected to be $0.112 \mu m^2$ and $0.047 \mu m^2$, respectively. This would potentially improve synaptic density by 2-3 orders of magnitude, dramatically improving the efficiency of the CMOL approach presented here. 

Research is also ongoing into the use of two-terminal nano-scale selector devices to replace the current 1T NMOS transistor \cite{Chen}. These two-terminal selectors would operate more similarly to a diode (or diffusive memristor) and could be sandwiched in series with the regular synaptic memristor at the same nano-wire crossing (named 1S1R). This would open the doors to true CMOL realizations, as originally envisioned by Likharev \cite{Likharev}, and to multiple layers of synaptic fabrics in 3D \cite{Linares2018}. Such 1S1R synapses could be fabricated with a pitch of about 100nm in a 22nm technology per layer, resulting in a synaptic density of 100 synapses per $\mu m^2$ (this is, $10^8$ synapses per mm$^2$)\footnote{The human cortex has about 2000cm$^2$ with 2$\times 10^{10}$ neurons and 10$^4$ synapses per neuron, resulting in 10$^9$ synapses per mm$^2$.}.

Another interesting undergoing research for removing the need of selectors, is the development of forming-free or born-on memristors \cite{ielmini_FF}. Memristor forming is the most critical operation demanding the need of a selector, as the forming voltage is typically higher than the required write voltage. However, if all memristors have a smaller forming than writing voltage and the latter is reasonable stable among memristors within a selector-less crossbar, then it is possible to rely on a $V_{write}/2$ scheme \cite{Radu} to safely address individual memristors for write or erase operations.

Another ongoing line of research into improving density and energy efficiency is the use of memristive synapses with analog weights \cite{Wang, Park, Hu, Le2019,Le2021, Giovinazzo,relaxation2016,ElisaArxiv2022}. Using analog weights for synapses can potentially reduce the number of synapses and neurons required to perform the same computations \cite{Yousefzadeh}. At the moment there are some published works reported in exploiting multi-level analog values per memristor, however as of today, for good separation, low resistances are required~\cite{Le2019,Le2021} and also there is a stochastic relaxation behaviour that takes several seconds after programming, thus slowing down the time efficiency of analog programming~\cite{relaxation2016,ElisaArxiv2022}.

Possible future developments include not only increasing the CMOL-core size and density by exploiting improved technologies, but also using these CMOL-cores as modules in large-scale single- or multi-chip systems where they can be re-configured and interconnected in a modular fashion to assemble larger neural layers or multi-layer computing systems. 

Finally, another interesting future outlook would be to replace the query-driven read-out by a native event-driven read-out. However, this would require the development of in-neuron circuitry for adapting and memorizing the indivdual firing thresholds for each neuron. Once reliable analog memristor memories with small incremental and stable updates become available, this would be a very interesting feature to add in a compact manner.

\section{Appendix}
\begin{figure}
\centering
\includegraphics[width=3.5in]{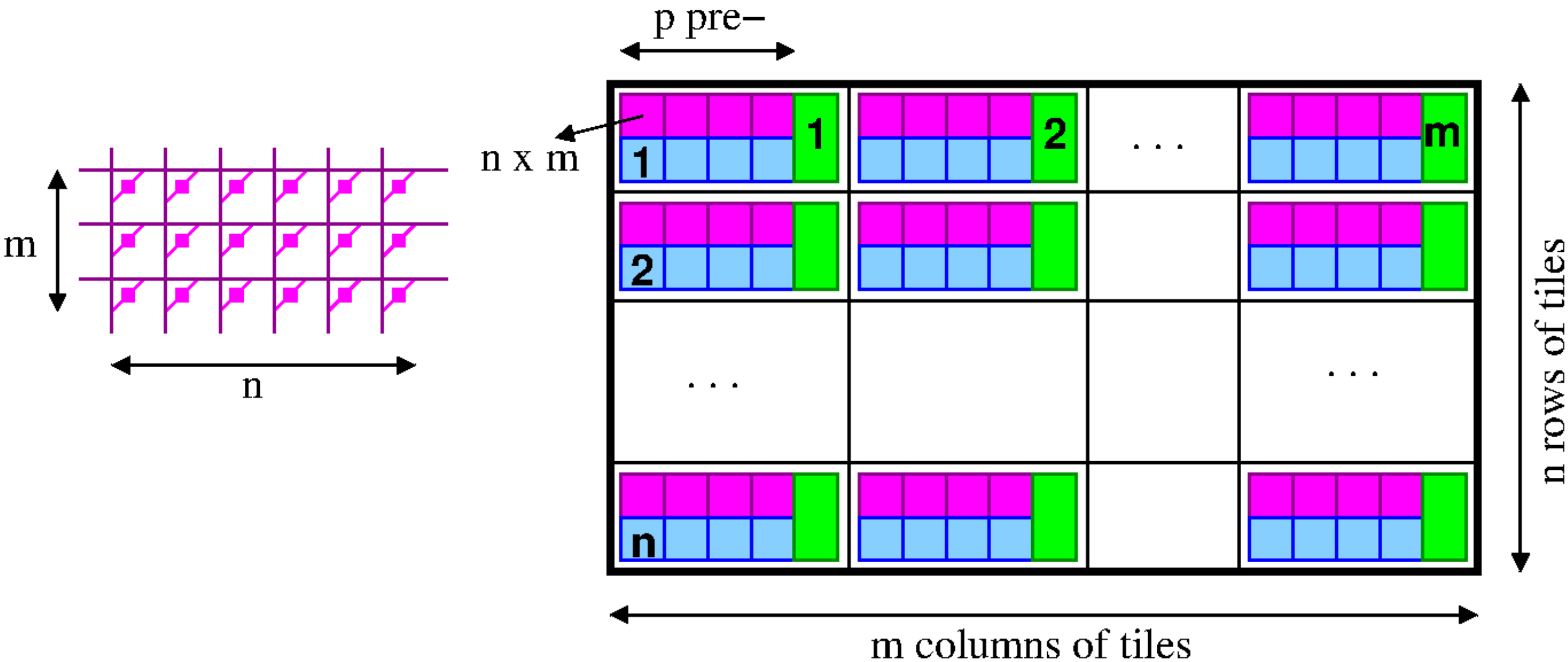}
\caption{Illustration of layout tiling with asymmetric crossbar.
}
\label{fig:Fig16}
\end{figure}

Here we illustrate the more generic case of asymmetric crossbars. Fig.~\ref{fig:Fig4} illustrates the case of a square crossbar with $M = 64$ pre-synaptic neurons, $N = 64$ post-synaptic neurons, and $N\times M = 4096$ synapses, but with $N=M$. In general, however, in multi-layer neural networks, the number of pre- and post-synaptic neurons is different. 
For this, let us consider that $M = p \times N$. Fig.~\ref{fig:Fig16} illustrates this case. There are $n$ rows of tiles and $m$ columns of tiles. But now each tile contains $p$ pre-synaptic neurons, one post-synaptic neuron, and $p$ synaptic sub-crossbars each of size $n \times m$. This way, the total number of post-synaptic neurons is $N= n \times m$, the total number of pre-synaptic neurons is $M = p \times n \times m$, and the total number of synapses is $(n \times m \times p) \times (m \times n) = M \times N$.



\newpage

 




\vfill

\end{document}